\documentclass[fleqn,usenatbib]{mnras}

\usepackage{newtxtext,newtxmath}

\usepackage[T1]{fontenc}

\DeclareRobustCommand{\VAN}[3]{#2}
\let\VANthebibliography\thebibliography
\def\thebibliography{\DeclareRobustCommand{\VAN}[3]{##3}\VANthebibliography}

\usepackage{pdflscape} 

\usepackage{graphicx}	
\usepackage{amsmath}	
\usepackage{enumitem}
\usepackage{color,soul}

\usepackage{orcidlink}

\graphicspath{ {} } 

\title[Validation of TOIs orbiting solar analogues]{Validation of TESS exoplanet candidates orbiting solar analogues in the all-sky PLATO input catalogue}

\author[G. Mantovan et al.]{Giacomo Mantovan$^{\orcidlink{0000-0002-6871-6131}}$,$^{1,2,3}$\thanks{E-mail: giacomo.mantovan@phd.unipd.it}
Marco Montalto$^{\orcidlink{0000-0002-7618-8308}}$,$^{4}$
Giampaolo Piotto,$^{1,2}$
Thomas G. Wilson$^{\orcidlink{0000-0001-8749-1962}}$,$^{3}$
\newauthor
Andrew Collier Cameron$^{\orcidlink{0000-0002-8863-7828}}$,$^{3}$
Fatemeh Zahra Majidi$^{\orcidlink{0000-0002-8407-5282}}$,$^{1,2}$
Luca Borsato$^{\orcidlink{0000-0003-0066-9268}}$,$^{2}$
Valentina Granata$^{\orcidlink{0000-0002-1425-4541}}$,$^{1,2}$
\newauthor
Valerio Nascimbeni$^{\orcidlink{0000-0001-9770-1214}}$$^{2}$
\\
$^{1}$Dipartimento di Fisica e Astronomia "Galileo Galilei", Università di Padova, Vicolo dell'Osservatorio 3, IT-35122, Padova, Italy\\
$^{2}$Istituto Nazionale di Astrofisica - Osservatorio Astronomico di Padova, Vicolo dell'Osservatorio 5, IT-35122, Padova, Italy
\\
$^{3}$Centre for Exoplanet Science, SUPA School of Physics and Astronomy, University of St Andrews, North Haugh, St Andrews KY16 9SS, UK
\\
$^{4}$Istituto Nazionale di Astrofisica - Osservatorio Astrofisico di Catania, Via S. Sofia 78, IT-95123, Catania, Italy}

\date{Accepted XXX. Received YYY; in original form ZZZ}

\pubyear{2021}

\begin{document}
\label{firstpage}
\pagerange{\pageref{firstpage}--\pageref{lastpage}}
\maketitle

\begin{abstract}
The Transiting Exoplanet Survey Satellite (\textit{TESS}) is focusing on relatively bright stars and has found thousands of planet candidates. However, mainly because of the low spatial resolution of its cameras ($\approx$ 21 arcsec/pixel), \textit{TESS} is expected to detect several false positives (FPs); hence, vetting needs to be done. Here, we present a follow-up program of \textit{TESS} candidates orbiting solar-analogue stars that are in the all-sky PLATO input catalogue. Using \textit{Gaia} photometry and astrometry we built an absolute colour-magnitude diagram and isolated solar-analogue candidates' hosts. We performed a probabilistic validation of each candidate using the \textsc{vespa} software and produced a prioritized list of objects that have the highest probability of being genuine transiting planets. Following this procedure, we eliminated the majority of FPs and statistically vetted 23 candidates. For this remaining set, we performed a stellar neighbourhood analysis using \textit{Gaia} Early Data Release 3 and centroid motion tests, greatly enhancing the on-target probability of 12 of them. We then used publicly available high-resolution imaging data to confirm their transit source and found five new, fully validated planets. For the remaining candidates, we propose on-off photometry to further refine the list of genuine candidates and prepare for the subsequent radial velocity follow-up. 

\end{abstract}




\section{Introduction}
The Transiting Exoplanet Survey Satellite (\textit{TESS}, \citealt{2014SPIE.9143E..20R}) is a NASA all-sky survey telescope designed to search for transiting exoplanets orbiting nearby stars. With its array of four ultra-wide-field cameras, \textit{TESS} has been delivering, since July 2018, both short-cadence photometry and target pixel file images on pre-selected targets, and full-frame images (FFIs) with a 30- or 10-minute cadence (during the nominal and extended mission, respectively). These data are downlinked to the ground, where they are then further analysed with transit-search pipelines developed by the Science Processing Operations centre (SPOC). This applies to short cadence images \citep{2016SPIE.9913E..3EJ} and, starting from sector 36, to some targets selected from the FFIs \citep{2020RNAAS...4..201C}, while every FFI is also analysed with the Quick-Look Pipeline (QLP, \citealt{2020RNAAS...4..204H}). The candidate planets found by the SPOC and QLP are then vetted\footnote{For details, see \url{https://heasarc.gsfc.nasa.gov/docs/tess/data-handling.html}, and \url{https://archive.stsci.edu/missions/tess/doc/EXP-TESS-ARC-ICD-TM-0014-Rev-F.pdf} (Twicken et al., 2020).} by the MIT branch of the \textit{TESS} Science office (TSO), and those candidates that survive are later defined as \textit{TESS} Objects of Interest (TOIs, \citealt{2021arXiv210312538G}). 

\textit{TESS} focuses on relatively bright, nearby stars and is finding thousands of transiting planet candidates. However, because of the low spatial resolution of its cameras ($\approx$ 21 arcsec/pixel), a percentage of objects initially identified as exoplanet candidates are expected to be false positives (FPs). In fact, the crowding of stars within the 1 arcmin$^2$ point spread function (PSF) of \textit{TESS} might cause two (or more) stars to appear merged into the \textit{TESS} time-series. Therefore, if an exoplanet orbits around a star that is blended with another in the \textit{TESS} images, then the transit signal in the light curve is diluted. If another star -- blended in the \textit{TESS} PSF -- is present in the same pixel, it could be the origin of the transit signal by either being an eclipsing binary or hosting a planet itself. 

Some FPs are identifiable using \textit{TESS} data alone, but the majority of them need further observations \citep{2015JATIS...1a4003R}. To avoid wasting observational time and optimize follow-up resources, it is possible to identify the most promising candidates through a quick and efficient probabilistic validation procedure, which aids in distinguishing between a planet and a FP from a particular transiting candidate \citep{2012ApJ...761....6M}. 

In this work, we present our probabilistic validation analysis of every TOI orbiting a solar-analogue target that is in the all-sky PLATO input catalogue \citep{2021A&A...653A..98M}, for which time-series or high-precision radial velocities follow-up observations are not yet available. We consider only candidates without follow-up observations available on the \textit{Exoplanet Follow-up Observing Program for TESS} (ExoFOP-TESS) website\footnote{Available at \url{https://exofop.ipac.caltech.edu/tess/}.}to provide an original analysis and avoid duplicated work. The software we use to perform such probabilistic validation is the \textsc{vespa} code, which is computationally efficient and publicly-available \citep{2012ApJ...761....6M}. By following this procedure, we are able to identify the majority of FP candidates. For the remaining set, we perform a stellar neighbourhood analysis using \textit{Gaia} Early Data Release 3 (\textit{Gaia} EDR3, \citealt{2021A&A...649A...1G}) and on-off photometry \citep{2009A&A...506..343D} to further refine the list of candidates and prepare for the subsequent radial velocity follow-up. As we will explain further in Section \ref{sec:definition}, throughout this work we label every candidate that passes only the \textsc{vespa} analysis as a `vetted' candidate; on the other hand, we refer to those that also pass the stellar neighbourhood analysis and meet a specific list of constraints as `statistically validated planets'. Furthermore, our work is a perfect case study of using \textsc{vespa} on PLATO data in the future, as the telescopes will have a similar spatial resolution \citep{2017SPIE10564E..05L}. This study could be the framework for future PLATO vetting.

In Section \ref{sec:methods} we briefly describe the methods we used to perform probabilistic validation analysis and stellar neighbourhood analysis; in Section \ref{sec:targ} we explain how we selected our sample of TOIs orbiting PLATO solar-analogue targets, while in Section \ref{sec:results} we show the results of our validation analysis, paying specific attention to the planetary size of the statistically-vetted targets we found. In Section \ref{sec:discussion} we discuss our results, provide suggestions for the follow-up observations, specify the nomenclature used, and call attention to the importance of performing a stellar neighbourhood analysis. Concluding remarks are in Section \ref{sec:conclusion}.


\section{Methods}
\label{sec:methods}
In this work, we performed the fully automated probabilistic validation procedure following \citet{2012ApJ...761....6M} \& \citet{2016ApJ...822...86M} for 158 TOIs orbiting solar-analogue stars that are in the all-sky PLATO input catalogue, v1.1 (asPIC1.1, \citealt{2021A&A...653A..98M}). The selection of these stars is described in Section \ref{sec:targ}, while in the following of this Section we describe the validation algorithms.

\subsection{VESPA}
The \textsc{vespa} code (\textit{Validation of Exoplanet Signals using a Probabilistic Algorithm}) is a publicly-available software package \citep{2012ApJ...761....6M} that models light curves of eclipsed stars as simple trapezoids parameterized by a depth $\delta$, a total duration $T$, and the transit shape parameter $T/\tau$, where $\tau$ is the \textit{ingress} (or \textit{egress}) duration, and simulates physically realistic populations of astrophysical FPs. 

Validating an exoplanet candidate is equivalent to demonstrating that the \textit{False Positive Probability} (FPP) is small enough to be considered negligible. \textsc{vespa} calculates the FPP as follows:
\begin{equation}
    {\rm FPP} = 1-{\rm Pr(planet|signal)}, \label{eq}
\end{equation}
 where 
\begin{equation}
    {\rm Pr(planet|signal)}= \frac{\mathcal{L}_{\rm TP}\pi_{\rm TP}}{\mathcal{L}_{\rm TP}\pi_{\rm TP}+\mathcal{L}_{\rm FP}\pi_{\rm FP}}
    \label{eq2}
\end{equation} defines the probability that there is a planet given the observed signal. In equation \ref{eq2}, $\mathcal{L}$ represents the Bayesian likelihood factor, which says how similar is the shape of the observed transit signal to the expected signal shape produced by the hypothesis (false positive or planet scenarios). The prior $\pi$ describes how intrinsically probable a priori is the existence of the hypothesized scenario. In particular, ${\rm TP}$ indicates a "true positive".

\textsc{vespa} supports the following hypotheses: 
\begin{itemize}[leftmargin=*]
    \item Eclipsing binary system in the background or foreground, blended within the photometric aperture of the target star (BEB);
    \item The target is a hierarchical-triple system where two of the components eclipse each other (HEB);
    \item The target star is an eclipsing binary (EB);
    \item Transiting planet (Pl)\footnote{\textsc{vespa} does not consider "blended transiting planet" FP scenarios \citep{2016ApJ...822...86M}.}.
\end{itemize} Furthermore, \textsc{vespa} supports double-period versions of each FP scenario. This is done to avoid, for example, the case in which an EB with twice the orbital period of the detected candidate, and with similar primary and secondary eclipse depths, is confused with a transiting planet. 

Briefly, the \textsc{vespa} validation procedure works in this way:
\begin{enumerate}[leftmargin=*]
    \item Simulation of a representative population for each hypothesis scenario listed above (fixing the period). Each population is made up of many different instances of that scenario;
    \item Calculation of the \textit{prior} ($\pi$) for each scenario, which is the product of three factors: the existence probability of the analysed scenario within the photometric aperture, the geometric probability of orbital alignment for which an eclipse is visible and the probability that the eclipse is able to mimic a transit \citep{2012ApJ...761....6M}; 
    \item Calculation of the \textit{likelihood} ($\mathcal{L}$) of the observed transit signal for each scenario, where \textsc{vespa} models the shape of the eclipse and fits it to the observed light curve. This is done through \textit{Markov Chain Monte Carlo} (MCMC) simulations;
    \item Combination of \textit{prior} and \textit{likelihood} to calculate the FPP of the transit signal (Equation \ref{eq}). If the FPP is $<$ 1 per cent, then the candidate can be considered as probabilistically vetted\footnote{It is crucial to note that, as further explained by \cite{2016ApJ...822...86M}, a vetted candidate requires to have a `probability > 99\% of being on the target star' to be fully `validated'. Therefore, as we explain in Section \ref{sec:definition}, we will label a candidate as fully `validated' only after proving this constraint. }. 
\end{enumerate}

We refer the reader to \citet{2012ApJ...761....6M} \& \citet{2016ApJ...822...86M} for a detailed description of the method.

\subsubsection{Data and constraints}
\label{data}
We fed \textsc{vespa} with the following stellar and planetary parameters:
\begin{itemize}[leftmargin=*]
    \item [-] equatorial coordinates, \textit{Gaia} photometric magnitudes, and parallax (see Section \ref{baye}) from \textit{Gaia} EDR3;
    \item [-] stellar effective temperature $T_{\rm eff}$, gravity log$g$, and metallicity [Fe/H]\footnote{We used $T_{\rm eff}$, log$g$ and [Fe/H] in the \textsc{vespa} calculation only if their values come from spectroscopy; otherwise, we avoided adding these input parameters.} from \textit{Mikulski Archive for Space Telescopes} (MAST);
    \item [-] mean stellar density $\rho$ in units of [{\rm g cm$^{-3}$}] and maximum extinction in the V band (\textit{maxAV}) from asPIC1.1;
    \item [-] planet to stellar radius $R_p/R_s$ from \textit{Exoplanet Follow-up Observing Program for TESS} (ExoFOP-TESS).
\end{itemize} We used the detrended and phase-folded time-series extracted from \textit{TESS} data available in the MAST portal (see Section \ref{phot}).

We defined the maximum angular distance (\textit{maxrad} in \textsc{vespa}) from the target star where a potential blending star might be, as the radius of the aperture ($r_{\rm circ}$) for circular aperture photometry. Otherwise, we assumed the area ($A$) covered by the \textit{TESS} aperture as circular and computed the radius as:
\begin{align}
\label{eqn:eqlabel}
\begin{split}
    r_{tess} &= \sqrt{A/\pi};
    \\
    A &= N_{\rm px} \times s^2,
\end{split}
\end{align} where $N_{\rm px}$ is the number of pixels within the \textit{TESS} aperture and $s$ is the \textit{TESS} platescale that is equal to 21 arcsec/pixel. As a safety margin, it is useful to add the \textit{TESS} PSF of 40 arcsec to both radii \citep{2018AJ....156..102S}, i.e., $\textit{maxrad} = r_{\rm circ} + 40$ arcsec and $\textit{maxrad} = r_{tess} + 40$ arcsec.

As described by \citet{2015ApJS..217...16R} and \citet{2016ApJ...822...86M}\footnote{\label{footnote_1}And following a tutorial of \textsc{vespa} available at \url{https://nexsci.caltech.edu/workshop/2018/VESPA_Tutorial.pdf}.}, we quantified the maximum depth of a potential secondary eclipse ($\delta_{\rm max}$, \textit{secthresh} in \textsc{vespa}). We ran a transit search in the TOI light curve and looked for the deepest signal allowed at phases outside the transit ($\delta_{\rm sec}$). We compute $\delta_{\rm max}$ as $\delta_{\rm max} = \delta_{\rm sec} + 3\sigma_{\rm sec}$, where $\sigma_{\rm sec}$ is the uncertainty associated with $\delta_{\rm sec}$.

Then, we inferred physical properties of the star (see Sec. \ref{sec:stellar}) given the photometric, spectroscopic, and observational constraints described above using the \textsc{Isochrones} package \citep{2015ascl.soft03010M} and finally computed the FPP with \textsc{vespa}.

\subsubsection{Bayesian evidence}
\label{baye}
As explained in \citet{2016ApJ...822...86M}, to start the validation procedure, all available constraints on the target star are used to condition a direct fit of a single- or multiple-star model to the MIST grid of stellar models (\citealt{2016ApJS..222....8D, 2016ApJ...823..102C, 2011ApJS..192....3P}). This fit is done using multi-modal nested sampling, implemented with \textsc{MultiNest}. Consequently, \textsc{Isochrones} produces posterior samplings of the physical properties of the host star, modelled as a single- or multiple-star system. To compute the FPP, \textsc{vespa} requires the physical properties of each stellar model (single, binary, and triple) to evaluate each different scenario. When multi-modal posteriors are sampled with the \textsc{MultiNest} tool, the Bayesian evidence \citep{2019OJAp....2E..10F} is also computed. This particular parameter allows us to understand the degree to which the data imply a given model \citep{KNUTH201550}. Therefore, it is usual to prefer the model that implies the greatest Bayesian evidence \citep{kass1995bayes}. 

In our validation procedure, we found out that inserting as input only the observed \textit{Gaia} photometric magnitudes -- instead of adding other photometric magnitudes -- produced the strongest Bayesian evidence. For this reason, we preferred to insert only the \textit{Gaia} G, BP, and RP magnitudes into the \textsc{vespa} input file.

\subsubsection{TESS photometry}
\label{phot}
\begin{figure}
	\includegraphics[width=\columnwidth]{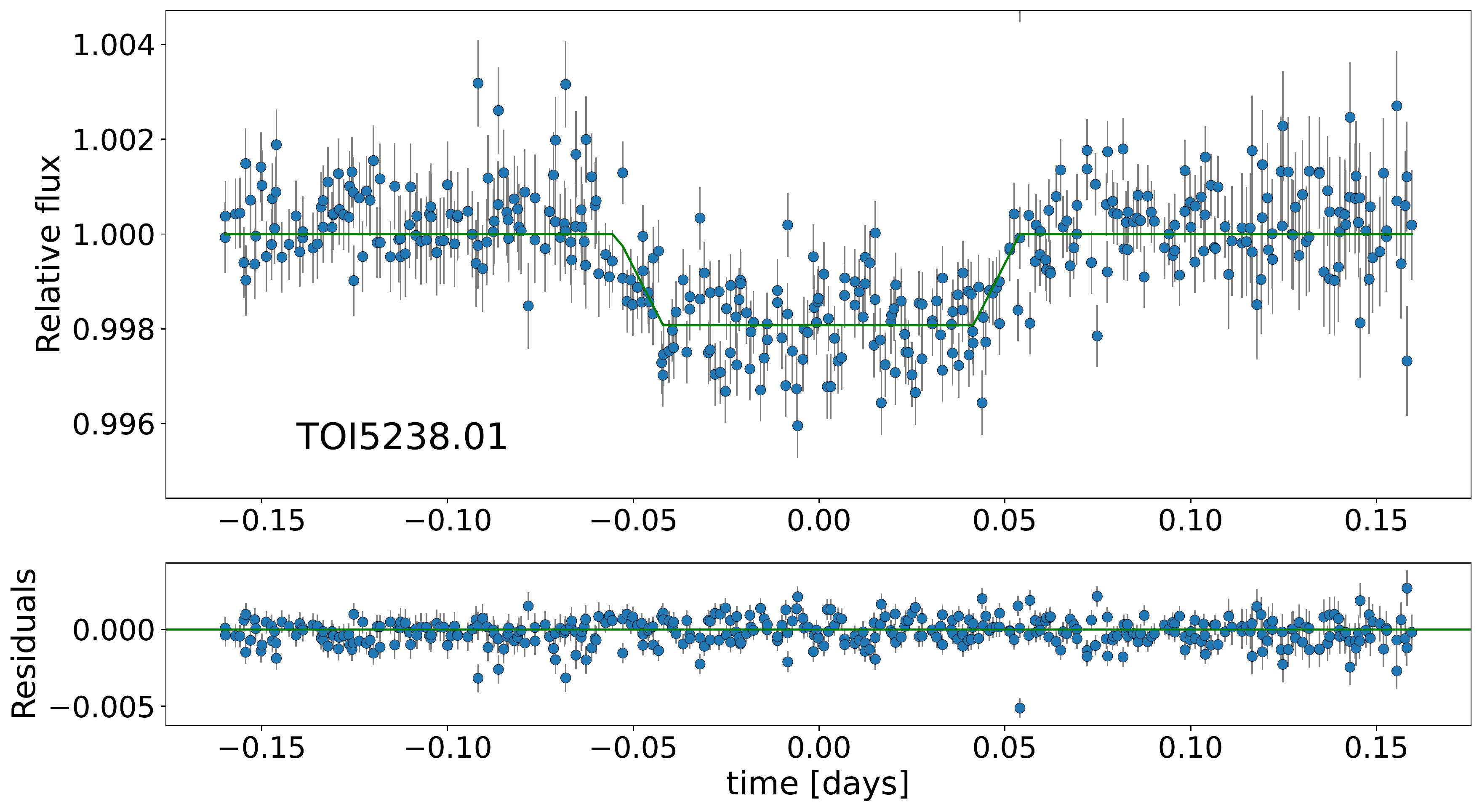}
    \caption{Detrended, normalized, and phase-folded light curve from the FFIs of TOI 5238.01. This \textit{TESS} candidate was observed in Sector 14, 15, 16, 21, 22, 23, and 41 and flagged as a planet candidate by the QLP on 2022 March 1. The time interval is centred on the mid-transit point and is limited to 1.5 times the transit duration on each side. The green line is a trapezoidal fitting model that is produced with the \textsc{vespa} software using the \textsc{batman} Python package \citep{2015PASP..127.1161K}. We plotted the residuals below the light curve.}
    \label{fig:qlp-fit}
\end{figure}

For our analysis of the \textit{TESS} light curves, we accessed the \textit{TESS} data by downloading the SPOC Presearch Data Conditioning Simple Aperture Photometry (PDCSAP) flux light curves (\citealt{2012PASP..124.1000S,2014PASP..126..100S}) for the short cadence candidates, which have been observed in multiple sectors and have been stitched together by the \textit{TESS} mission in the so-called \textit{Data Validation Time Series} files. These files can be found in the MAST portal. When instead we deal with TOIs from FFIs, we downloaded the QLP normalized light curves detrended by splines (KSPSAP). In cases where QLP multisectors observations had been available, we stitched together each light curve and then we performed the phase-folding procedure (Fig. \ref{fig:qlp-fit}).

\subsection{Stellar neighbourhood analysis}
\label{stellar-neighbourhood}
To further understand the real nature of a \textit{TESS} candidate, it is necessary to accurately analyse its stellar neighbourhood, to understand if possible contaminating stars are present.

Thanks to \textit{Gaia} photometry \citep{2021A&A...649A...1G}, we can check whether any neighbourhood star is able to generate a flux in the aperture that corresponds at least to the observed flux variation. When this is not the case, we can exclude each \textit{Gaia} source to be a blended eclipsing binary and greatly enhance the probability that the detected signal is coming from the target star. Consequently, we developed a custom pipeline (further explained at the end of this Section and in Section \ref{subsec:diluted} \& \ref{k-param}) to compute which neighbourhood stars could reproduce the observed transit signal. When none of them can, we changed the value of the \textit{maxrad} constraint within the \textsc{vespa} input data and proceeded with the analysis. In this case, we considered the spatial resolution of \textit{Gaia} EDR3 -- that can resolve close pairs of stars at 1.5 arcseconds separation -- as the minimum value for the \textit{maxrad} constraint. It is important to note that this is a peculiar situation, which only happened to $\sim$ 1/10 of our targets.

Therefore, it is often necessary to perform additional photometric follow-up observations to confirm the source of the signal. We can apply the so-called \textit{seeing-limited on-off photometry} technique \citep{2009A&A...506..343D}, which consists of the flux measurement -- with ground-based imagers -- of the target star and neighbour stars within, for example, 3.5 arcmins (10x10 \textit{TESS} pixels, \citealt{2019AJ....158..138S}) during the predicted on- and off-transit phases. Thanks to the much higher angular resolution obtainable with some ground-based instruments compared to \textit{TESS}, we can either confirm or discard that the detected signal is due to a genuine exoplanet candidate orbiting a given target star. Using these high angular resolution imagers, we can first resolve most cases of stars that appear merged into the time-series obtained by \textit{TESS}; then -- depending on the photometric precision of the instrument and the transit depth -- we can either: 
\begin{enumerate}[leftmargin=*]
    \item detect the source of the signal and verify if it does not exhibit luminosity variations that are sufficiently strong to cause a false alarm (when we have high photometric precision and/or a deep transit depth); or 
    \item focus on the photometry of the neighbourhood stars and verify if none of them can reproduce the discovery signal (when we have low photometric precision and/or a shallow transit depth).
\end{enumerate} Both strategies are able to detect the source of the signal. It is necessary to note that for the success of this technique, it is crucial to take into account the \textit{TESS} ephemeris uncertainty (Epoch, Period, and Duration) and perfectly plan the on-off observation windows. In fact, the accumulation of the uncertainty over time shortens the window length in which we can precisely collect the off- and, especially, on-transit phases. Thanks to the multi-year \textit{TESS} observations, the ephemeris uncertainty is often less of an issue. Nonetheless, it is important to collect on-off photometry as close as possible to the last \textit{TESS} observation to avoid the accumulation of uncertainties. Therefore, when we have many data points and ephemeris uncertainties are small, we should shorten by 3$\sigma$ both sides of the window length of the on-transit phase, where $\sigma$ takes into account the ephemeris uncertainties and the ingress/egress duration. In addition to that, it is essential to perform this follow-up with an observation band similar to \textit{TESS}, such as the Cousins $I_{\rm c}$ or the Sloan $i^{\prime}$, to avoid the potential obtainment of a transit depth different than expected (see Sec. \ref{sec:on-off}).

Both in the on-off technique and in the photometric analysis using \textit{Gaia} EDR3, we computed the expected magnitude variation that any neighbourhood star have to generate to reproduce the transit signal. Specifically, we followed \citet{2009A&A...506..343D} and found that a transit signal originates from a neighbour star $\rm c$ if this star is able to reproduce the discovery signal ($s$):
\begin{equation}
    (\Delta F/F)_{\rm s} = \frac{k_{\rm c} \Delta F_{\rm c}}{k_{\rm t} F_{\rm t}+ \sum k_{\rm i} F_{\rm i}},
    \label{eq:signal}
\end{equation} where $\rm t$ and $\rm i$ (with $\rm c \in i$) stand for \textit{target} and \textit{contaminants} respectively, while $k$ is the \textit{fraction of light of the stellar PSF which falls into the given photometric aperture}. The aim of this photometric follow-up is then to falsify the equation \ref{eq:signal} for each neighbour star. When we are analysing a contaminant star, we can rewrite equation \ref{eq:signal} as follows:
\begin{equation}
    \frac{\Delta F_{\rm c}}{F_{\rm c}} = \left ( \frac{k_{\rm t} F_{\rm t}+ \sum k_{\rm i} F_{\rm i}}{k_{\rm c} F_{\rm c}}\right )(\Delta F/F)_{\rm s},
\end{equation} which in magnitude notation becomes:
\begin{equation}
    \Delta m_{\rm c} = -2.5 \log \left(1-(\Delta F/F)_{\rm s}\left(\frac{k_{\rm t}10^{-0.4 m_{\rm t}}+\sum k_{\rm i}10^{-0.4 m_{\rm i}}}{k_{\rm c}10^{-0.4 m_{\rm c}}}\right)\right).
   \label{eq:magexp}
\end{equation} The argument of the logarithm must be positive. This means that each contaminating star needs to generate a flux in the aperture that corresponds at least to the observed flux variation. If none of them can pass this threshold, we can rule out each resolved neighbourhood star as the source of the transit signal without taking a photometric observation to evaluate $\Delta m_{\rm c}$. In this specific situation, we can already move on to high-resolution imaging and precision radial velocity observations; otherwise, we require ground-based photometric observations to perform the on-off follow-up. 

\subsubsection{Diluted discovery signal}
\label{subsec:diluted}
In our procedure, the discovery signal in equation \ref{eq:signal} must be the one coming from simple aperture photometry. This is important because we need to conserve the possible stellar contamination coming from neighbour stars to subsequently correct it with our pipeline. However, the transit depth of a TOI -- provided by the TESS team -- comes from a PDCSAP \footnote{or KSPSAP, if the specific TOI has been identified with the QLP pipeline.} flux light curve (hereafter, $\delta_{\rm PDCSAP}$), which is already corrected for the crowding contamination from known neighbour stars \citep{2021arXiv210312538G}. A similar amount of flux correction is collected into the TIC \textit{contamination ratio} parameter \citep{2019AJ....158..138S}, which is defined as the nominal flux from the contaminants divided by the flux from the source. The contaminants have been searched for within 10 \textit{TESS} pixels, and the contaminating flux has been calculated within a radius that depends on the target's Tmag. Using this parameter, we can therefore recover, in a simplistic way, the diluted transit depth as follows:
\begin{equation}
    (\Delta F/F)_{\rm s} = \frac{\delta_{\rm PDCSAP}}{(1+{\rm CR})} 
\end{equation} where CR is the \textit{contamination ratio}. 

We recovered the diluted transit depth and performed a custom correction for stellar dilution for two reasons. Firstly, there are known cases where the QLP planet radius has been inaccurate relative to uncontaminated ground-based observations. This inaccuracy has often turned out to be linked to the QLP deblending method, which is based on the \textit{TESS} magnitude estimates from the TIC \citep{2020RNAAS...4..204H}. Moreover, the QLP deblending method effectively deblends the light curve from contamination by an additional star inside the aperture \citep{2021arXiv210312538G}, whereas we deblend the transit depth from contamination by any star whose flux falls inside the aperture. Secondly, SPOC simulates the contaminating flux in the field around the target star from the full TICv7 catalogue for sectors 1-13 and TICv8 for sectors 14 onwards \citep{2021arXiv210312538G}, and both use the \textit{Gaia} DR2 catalogue. In our pipeline, we used instead stellar parameters from TICv8.2 and included parameters from the \textit{Gaia} EDR3 catalogue. As we will specify in Section \ref{k-param}, we have used the \textit{TESS} photometric band to correct for stellar contamination. In particular, we obtained the \textit{TESS} magnitude of a \textit{Gaia} star by cross-matching TIC and \textit{Gaia} catalogues through the \textit{Gaia} ID of the star.

\subsubsection{The k parameter}
\label{k-param}
The $k$ parameter modifies the expected magnitude difference that is required to reproduce the transit signal. Its value depends on the selected pixels and on the exact position of the given star in the \textit{TESS} aperture. In fact, the PSF of the telescope causes the light from the target to fall onto several different pixels. The photometric aperture used to extract the light curve of a short-cadence TOI can be found inside a \textit{Target Pixel File} (TPF) object, which is an ensemble of images taken for each observed cadence. Differently, for long-cadence TOIs found with the QLP, the \textit{TESS} aperture is circular and its optimal radius is given by the QLP itself. 

Thanks to the \textit{Pixel Response Function} (PRF) provided by the \textit{TESS} mission, we can determine in which pixels the light from the target falls. In detail, the PRF is a model that describes the image of a point source and how it varies depending on where it lands on the detector. Its shape comes from a combination of  the optical point spread function, jitter during observations, and intra-pixel location of where the light lands\footnote{For details, see \url{https://archive.stsci.edu/files/live/sites/mast/files/home/missions-and-data/active-missions/tess/_documents/TESS_Instrument_Handbook_v0.1.pdf}.}. The PRF images span 13x13 physical \textit{TESS} CCD pixels and have 9x9 intra-pixel samples per each pixel. This ensures some pixel precision without having to interpolate. The available PRF files change among different \textit{TESS} sectors, cameras, and CCDs.

The entire procedure for evaluating the $k$ parameter for an individual star (whether it is the target or a contaminant) can be described as follows:
\begin{itemize}[leftmargin=*]
    \item We extract the PRF at the exact pixel location of the star, where the total flux is determined using the star's \textit{TESS} magnitude; 
    \item We evaluate the shift between the centre of the aperture and that of the PRF (i.e., the separation between the centre of the aperture and star location); 
    \item We calculate the exact contribution of the flux that falls into the aperture. When the aperture is circular, we use the implementation of the \textsc{photutils} package \citep{larry-bradley-2021-5525286}. When dealing with TPF apertures (made by multiple square pixels), we calculate four weights to consider the displacement between the centre of an aperture pixel and that of a PRF pixel. This displacement causes an aperture pixel to overlap with up to a maximum of four PRF pixels; therefore, to calculate the flux contribution from a single aperture pixel, we need to consider the weighted contribution of each of these four PRF pixels;
    \item We divide the contribution of flux that falls into the aperture by the total flux of the star. The result obtained is the $k$ parameter.
\end{itemize}
\subsubsection{Undiluted radius}
\label{sec:undiluted}
The analysis of the neighbour stars allows us to evaluate how much the stellar dilution affects the candidate's transit depth -- and thus its radius -- and whether this value remains consistent with a planetary object. The equation of the new transit depth is as follows:
\begin{equation}
    \frac{\Delta F_{\rm t}}{F_{\rm t}} = \left ( \frac{k_{\rm t} F_{\rm t}+ \sum k_{\rm i} F_{\rm i}}{k_{\rm t} F_{\rm t}}\right )(\Delta F/F)_{\rm s},
\end{equation} with the same notation used in eq. \ref{eq:signal}, while the candidate's new radius ($R_{\rm p}$) can therefore be estimated with the following equation: 
\begin{equation}
    R_{\rm p} \approx R_{\star}\sqrt{\frac{\Delta F_{\rm t}}{F_{\rm t}}}.
    \label{eq:undiluted}
\end{equation}

\subsection{Centroid Motion}
\label{centroid}
In addition to the on-off photometry technique, we performed another verification test to recognise the presence of contaminating stars. In particular, we have exploited the so-called \textit{centroid motion} test, which monitors the shift in the position of the photometric centroid during a transit event and verifies whether the corresponding motion is pointing away from the target. With this test, we further seek to determine the location of the transit source and to discard blended eclipsing binary sources. 

In the specific case where a TOI was identified by the SPOC, we took the result of the centroid motion test carried out by the \textit{TESS} mission -- that can be found within a \textit{TESS} \textit{Data Validation Report} file. When this is not the case, we followed the procedure described in \citet{2020MNRAS.498.1726M} to perform the centroid test, and then we applied the suggested constraints to determine whether a candidate has passed the test. These constraints include the probability of correct source identification $\rm P_\eta$, the probability of correct source association $\rm P_D$, and the Mahalanobis distance \citep{mahalanobis1936generalized}.

\section{Targets selection}
\label{sec:targ}
To select the candidates orbiting solar-analogue stars, we built an intrinsic colour-magnitude diagram in the \textit{Gaia} bands, correcting the photometry for distance modulus, extinction, and reddening. From this diagram, it is possible to extrapolate all stars belonging to a certain spectral class. We focus on solar-analogue stars because of the scientific importance of discovering planets around `Solar twins' to carry out future atmospheric follow-up, and to obtain statistical information about exoplanet systems, whose characteristics may be similar to our planet, the only one known to host life. 

\subsection{Intrinsic colour-magnitude diagram}
We developed a custom pipeline that takes into account the entire list of TOIs and then cross-matches it with the MAST, which also includes \textit{Gaia} data. Then, we cross-matched the same list with the asPIC1.1 \citep{2021A&A...653A..98M} and used its corrected \textit{Gaia} DR2 photometry to build an intrinsic colour-magnitude diagram.

\subsubsection{The all-sky PLATO input catalogue}

\begin{figure}
	\includegraphics[width=\columnwidth]{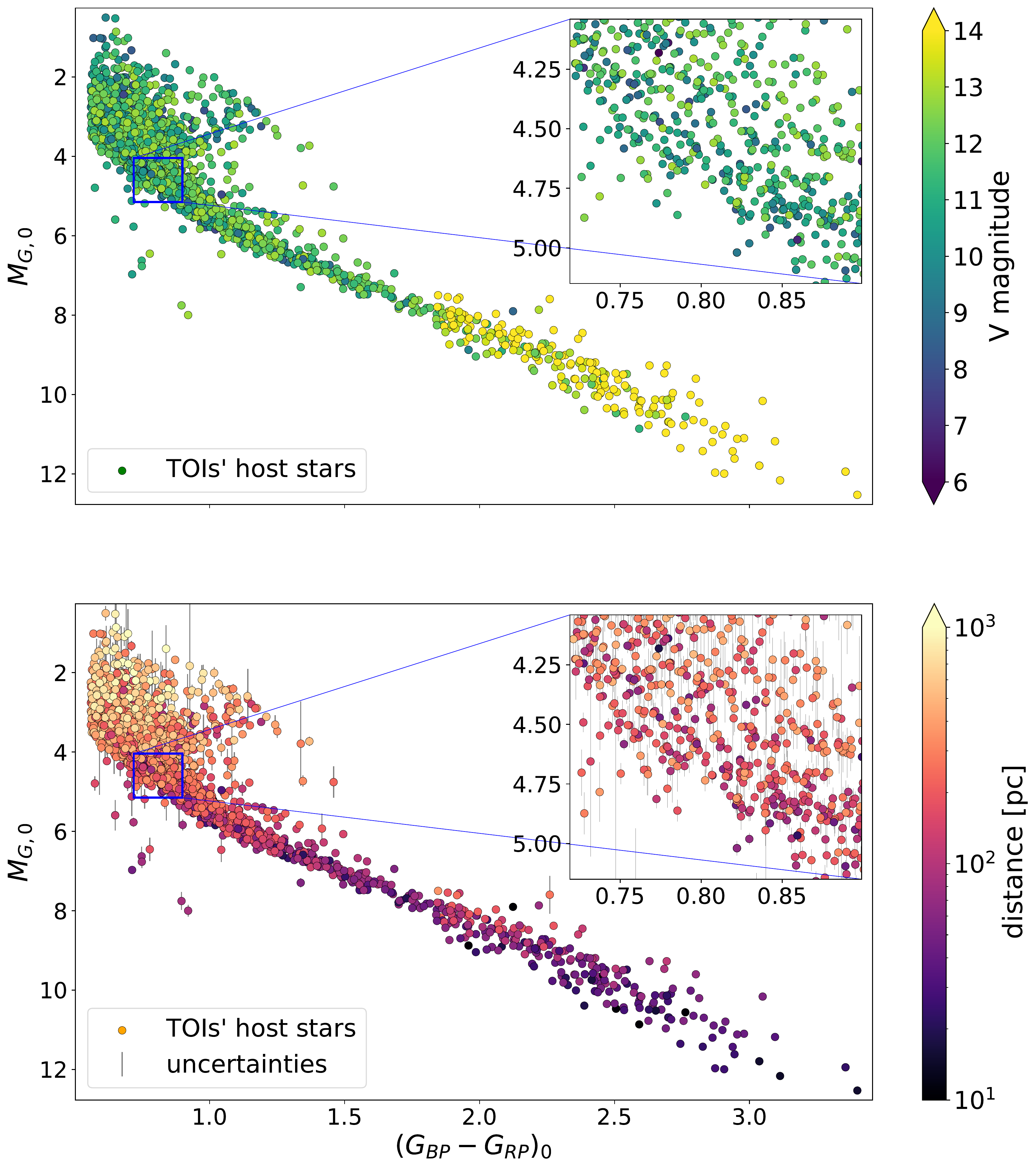}
    \caption{Intrinsic colour-magnitude diagram in the \textit{Gaia} bands. Each dot represents a specific TOI whose host star is inside the asPIC1.1 input catalogue, while either the apparent V magnitude (\textit{top} panel) or the stellar distance (\textit{bottom} panel) is colour-coded. In the \textit{bottom} panel, we added star-by-star uncertainties for the absolute, intrinsic $G_0$ magnitude. We also plotted our selected solar-analogue stars (see Sec. \ref{sec:selection}) within the insets.}
    \label{fig:colour-mag-mix}
\end{figure}

The intrinsic (i.e., reddening and extinction-free), absolute colour-magnitude diagram is presented in Figure \ref{fig:colour-mag-mix}. In the Figure, the V magnitude (\textit{upper} panel) and the stellar distance  (\textit{lower} panel) are colour-coded. This colour-magnitude diagram represents stars across all the range of spectral types isolated in the asPIC1.1 (FGK and M dwarfs and subgiant stars), where for FGK stars V $\leq$ 13 and for M dwarfs V $\leq$ 16. This completeness is important to highlight -- especially across the G spectral class -- since it ensures that we did not apply any significant bias in our selection. 

We found that in a magnitude limited sample, bluer stars -- being intrinsically more luminous -- tend to be located at larger distances than redder stars along the main sequence, which are instead found closer to the observer. Then the larger distance implies a larger distance, reddening, and extinction uncertainty. This also explains both the increase in colour and magnitude uncertainty (Fig. \ref{fig:colour-mag-mix}, \textit{lower} panel). Furthermore, the magnitude (and colour) uncertainty tends to increase towards bluer stars. Since interstellar extinction is inversely proportional to the wavelength \citep[Whitford's Law of Interstellar Extinction, ][]{1958AJ.....63..201W}, bluer stars tend to be more extincted, and hence the magnitude uncertainty increases. The colour uncertainty has the same tendency.


\subsection{Stellar sample}
\label{sec:selection}
At this stage, every TOI's host star within the asPIC1.1 is ready for selection. The data is continuously updated and at the time of writing it, 2022 May 16, the number of TOIs discovered is 5637, of which 2842 are included in the asPIC1.1. We decided to use Mamajek's table \citep{2013ApJS..208....9P} for these TOIs' host stars. This table provides average colours and magnitudes (in different pass bands) for each spectral class and hence allows us to do a selection based on these average values. We used the photometric magnitudes of the second\footnote{Although \textit{Gaia} EDR3 is now available, we used \textit{Gaia} DR2 in our stellar selection because both the asPIC1.1 and Mamajek's table are based on this \textit{Gaia} release.} \textit{Gaia} data release (corrected for extinction). We chose the stellar classes from F9V to G8V, providing almost seven hundred target stars to be analysed. The choice to use this range of stellar subclasses is arbitrary, but motivated by the colour and magnitude parameters of its two extremities, which are almost equally separated from the parameters of the Sun. Furthermore, both the F9V and the G8V subclasses differs by about $\pm$ 300K from the effective temperature of the Sun. Expanding the range at each end with only one subclass would imply an expansion into the effective temperature range of $\pm$ 120K, which is equivalent to a total range expansion of $\approx$ 40\%. After this selection, we considered only TOIs currently defined by the TESS Follow-up Observing Program (TFOP) working group as \textit{Planet Candidate} (PC) or whose definition\footnote{We extracted the dispositions from the `TFOPWG Disposition' entry available on the ExoFOP-TESS website.} is still absent. We also excluded from our analysis each TOI for which time-series or high-precision RV follow-up observations were already available and all TOIs currently under investigation by the ExoFop-TESS website. In total, after discarding single-transit candidates, 158 TOIs survived within our selection.

\section{Results}
\label{sec:results}
Here we report the result of the validation procedure for 158 TOIs orbiting solar-analogue stars analysed using the \textsc{vespa} code. In the following subsections, we present the statistical outputs coming from this calculation. 

\subsection{Stellar Parameters}
\label{sec:stellar}
The \textsc{vespa} code relies on the \textsc{Isochrones} package to infer physical properties of a \textit{TESS} star. \textsc{Isochrones} uses a nested sampling scheme given photometric, spectroscopic, and other observational constraints (see Section \ref{data}). Stellar properties are crucial for estimating the FPP \citep{2017ApJ...847L..18S}. If the \textsc{Isochrones} estimates agree with literature values, the resulting FPP becomes more reliable. 

We compared the stellar parameters simulated in this work with those determined for asPIC1.1 by inspecting the difference in their values:
\begin{equation}
    \Delta x_i = x_{i,{\rm vespa}} - x_{i,{\rm PIC}},
\end{equation} while its uncertainty is:
\begin{equation}
    \sigma_{\Delta x_i} = \sqrt{\sigma_{i,{\rm vespa}}^2+\sigma_{i,{\rm PIC}}^2},
\end{equation} where $x_i$ and $\sigma_i$ are the value and standard deviation of a given stellar parameter $i$ respectively, with $i$ = \{mass, radius, $T_{\rm eff}$, distance\}.

To perform the comparison, we used stellar parameters from the \textsc{Isochrones} single-star fit when the planetary or BEB scenario was the most likely; otherwise, we used those from either the double- (EB scenario) or triple-star model (HEB scenario). We have always used the stellar parameters of the primary star, both in the case of a single- and a multiple-star system. When \textsc{vespa} simulates the BEB scenario, it does not use stellar parameters from the \textsc{Isochrones} star models; instead, it performs a TRILEGAL simulation \citep{2005A&A...436..895G} to generate a population of eclipsing binary stars in the neighbourhood of the \textit{TESS} target under examination. The source of the transit signal becomes one of these neighbourhood stars; hence, the simulated stellar parameters that have been generated are different from those of our selected target star. However, we aim to verify the \textsc{vespa}-simulated stellar parameters of our \textit{TESS} star specifically selected and not one of its neighbour stars. Therefore, we made use of stellar parameters from the \textsc{Isochrones} single-star model that have been used to evaluate the planetary scenario, regardless of whether \textsc{vespa} identified the BEB-scenario to be most likely.

\begin{figure*}
	\includegraphics[width=\textwidth]{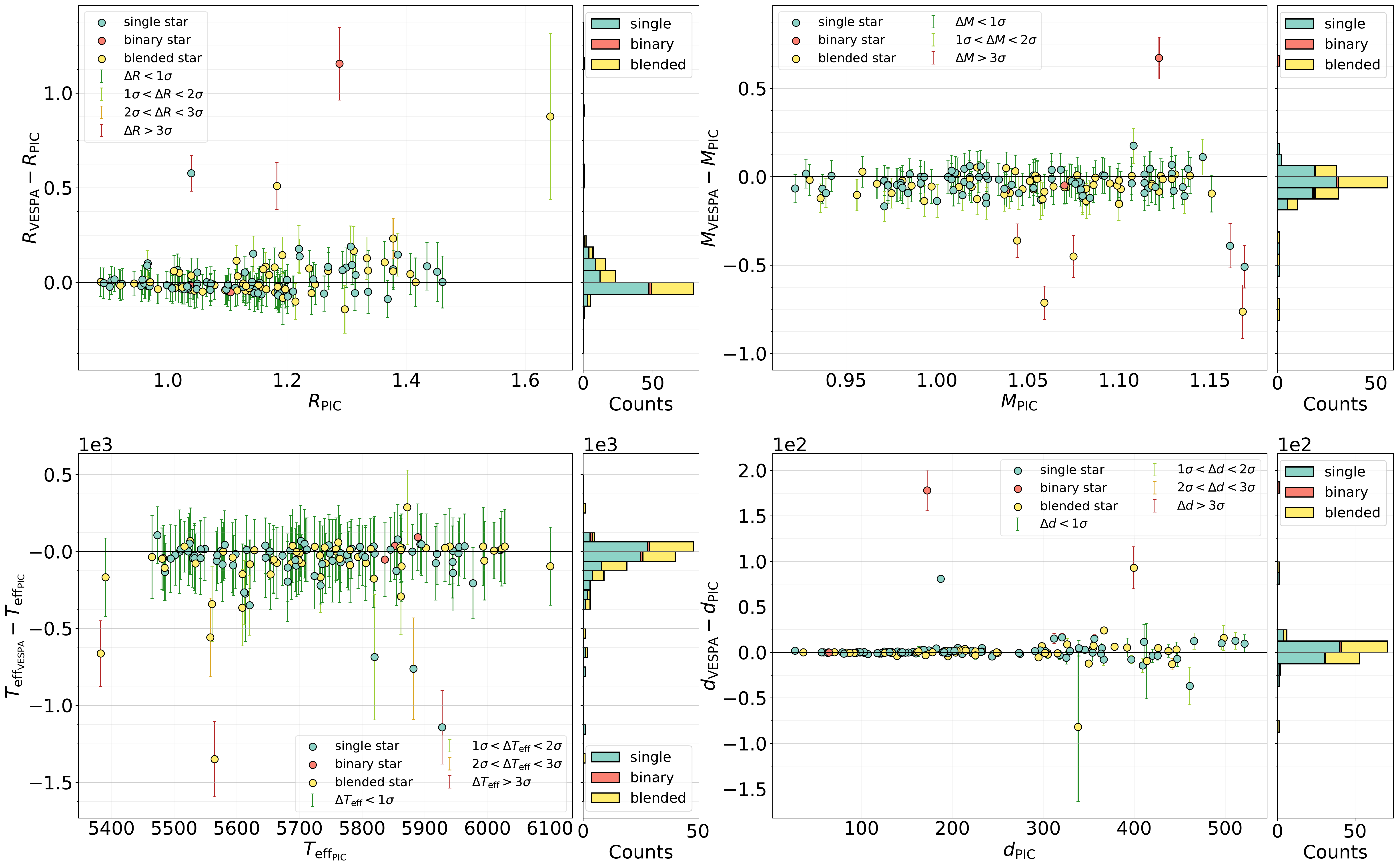}
    \caption{Level of agreement between \textsc{vespa}-simulated and asPIC1.1 stellar parameters. Each panel shows $x_{i,{\rm vespa}} - x_{i,{\rm PIC}}$ versus $x_{i,{\rm PIC}}$, where $x$ is the value of a given stellar parameter $i$, with $i$ = \{radius, mass, $T_{\rm eff}$, distance\}. The \textsc{vespa} most likely stellar scenario (single, binary, or blended star) and the uncertainty on $\Delta x_i$ are colour-coded. The latter has four different colours depending on the size of the uncertainty compared to its $\Delta x_i$. The histogram to the right of each panel plots the distribution of each value.}
    \label{fig:stellar}
\end{figure*}

Figure \ref{fig:stellar} shows the difference $\Delta x_i$ between the two measures versus the asPIC1.1 stellar parameter $x_{i,{\rm PIC}}$. We omitted stars with $|\Delta x_i| > 3\sigma_{\Delta x_i}$ from our analysis.

As we can see in Fig. \ref{fig:stellar}, we found that almost all \textsc{vespa}-simulated host stars have parameters in agreement with those estimated in the asPIC1.1 input catalogue; and, aside from four particular `single' star cases (see below), for every star outside $3\sigma$ confidence interval \textsc{vespa} found the BEB (or EB) scenario to be the most likely. We also noted the presence of a small systematic offset on stellar masses, which is probably due to the different empiric relationships or models adopted by asPIC1.1 and \textsc{Isochrones}. By the analysis of the individual targets, we see that:
\begin{itemize}[leftmargin=*]
    \item the candidates we statistically vetted (see Sect. \ref{valid}) orbit a star whose simulated parameters agree with the asPIC1.1 ones; 
    \item stars with a stellar parameter $i$ with $|\Delta x_i| > 3\sigma_{\Delta x_i}$ usually also have one (or more) other stellar parameters that follow this characteristic, which means that nine target stars had to be discarded from our analysis;
    \item stars outside $3\sigma$ confidence interval almost often had large \textit{maxAV} and/or $\Delta \rho / \rho$ values in the \textsc{vespa} input files. This was the case for three of the four `single' stars outside $3\sigma$ confidence interval. The large \textit{maxAV} value in the \textsc{vespa} input files -- that we noticed being quite often underestimated in the consecutive \textsc{vespa} simulation of these stars -- could explain both the differences in stellar $T_{\rm eff}$ and $M$. On the other hand, the large $\Delta \rho / \rho$ values could explain some of the stars with large \textsc{vespa}-simulated stellar radius $R$ and mass $M$; 
    \item the remaining unexplained `single' star was modelled by \textsc{Isochrones} but the resulting fit led to erroneous posterior stellar parameters. 
\end{itemize}

\subsection{False Positive Probability}
\begin{figure}
	\includegraphics[width=\columnwidth]{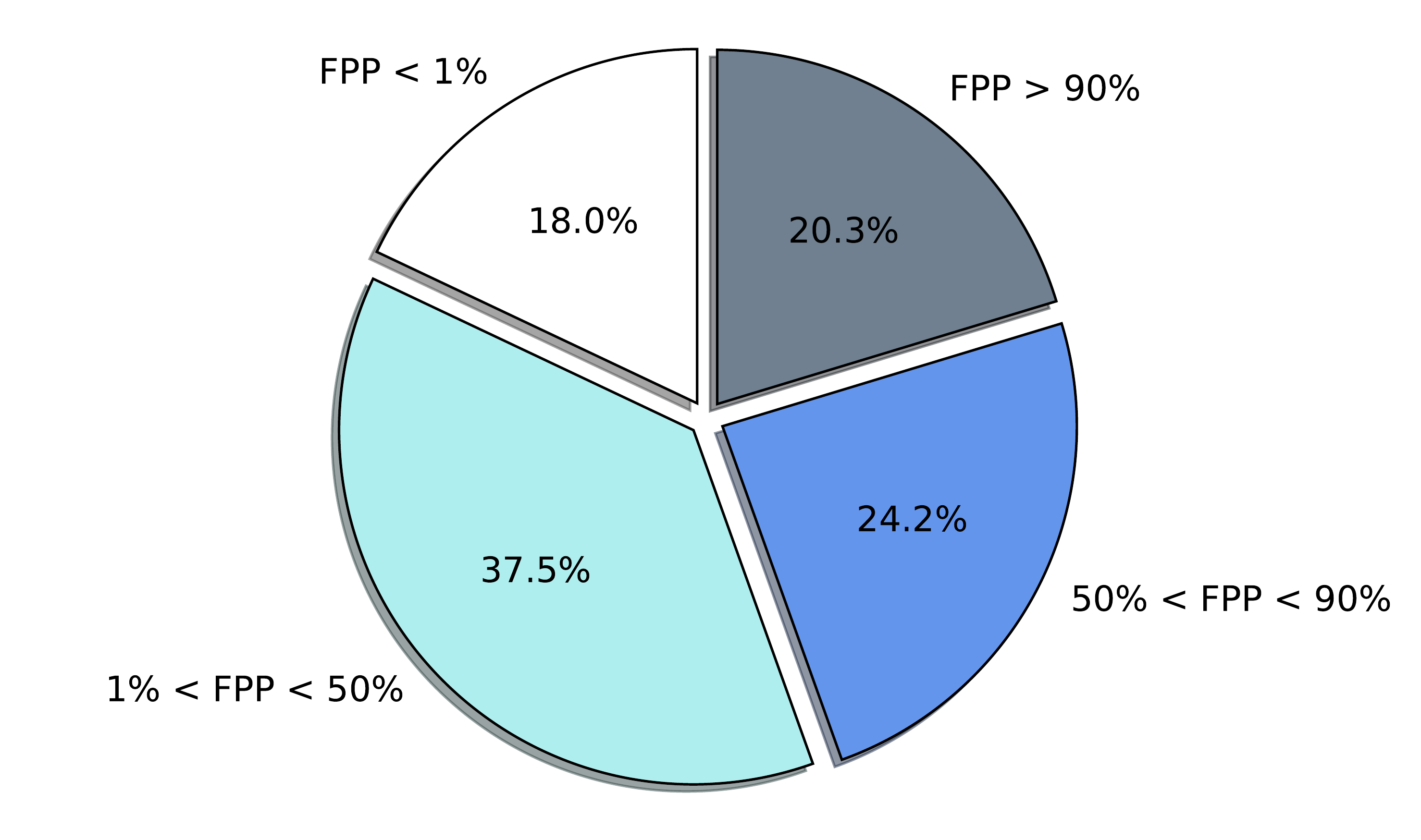}
    \caption{Distribution of the false positive probability for the 128 TOIs orbiting solar-analogue stars analysed. The white slice represents the number of candidates who are likely transiting planets. The light-blue one represents the number of those with a probability of 1\%<FPP<50\%, while the blue slice represents those with a probability of 50\%<FPP<90\%. The grey slice represents those that most likely are false positives.}
    \label{fig:fpp-pie}
\end{figure}

\begin{figure}
	\includegraphics[width=\columnwidth]{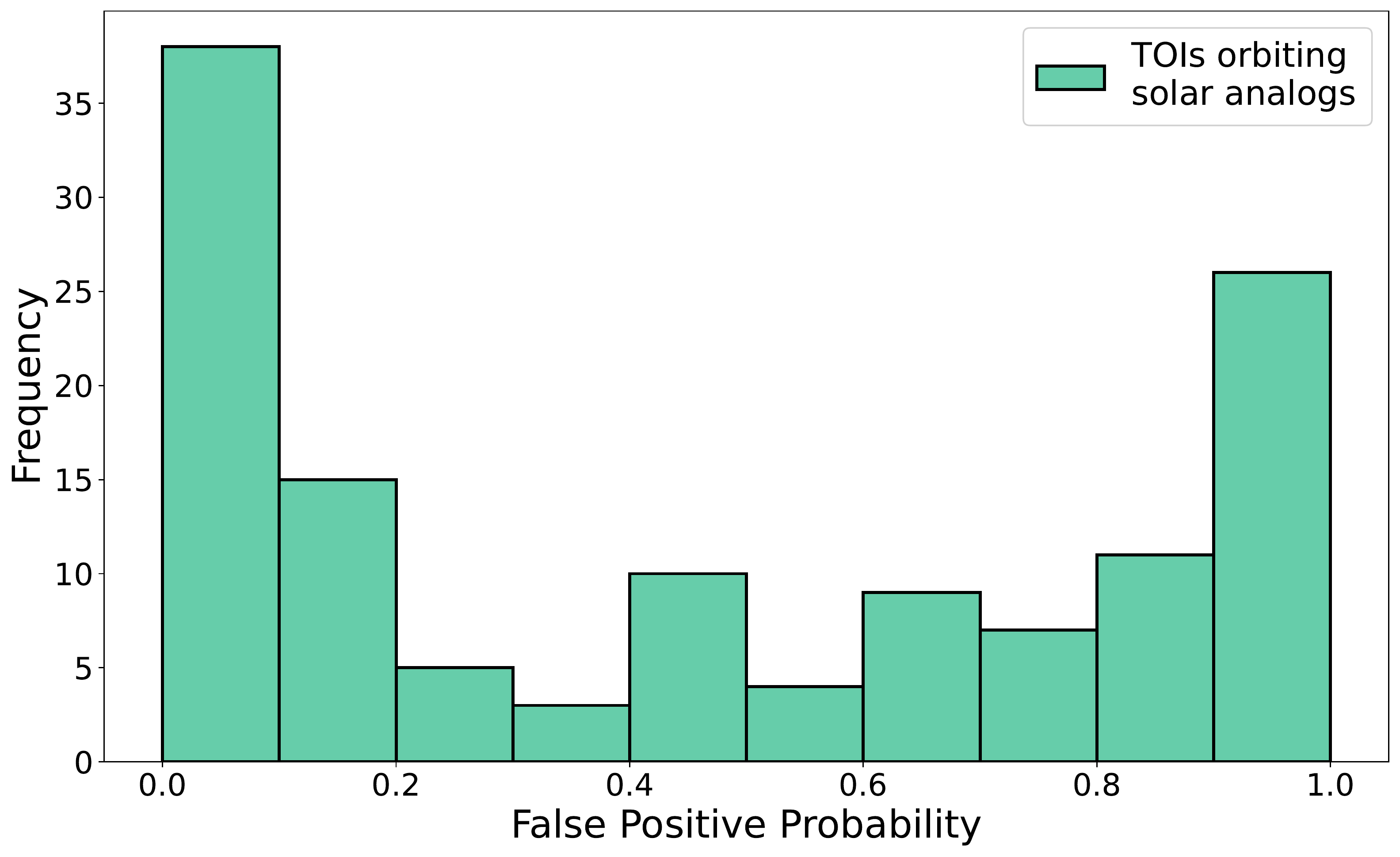}
    \caption{False positive probability distribution for the 128 TOIs orbiting solar-analogue stars analysed. We subdivided the FPP values from 0 to 1 in 10 bins.}
    \label{fig:isto}
\end{figure}

\begin{figure}
	\includegraphics[width=\columnwidth]{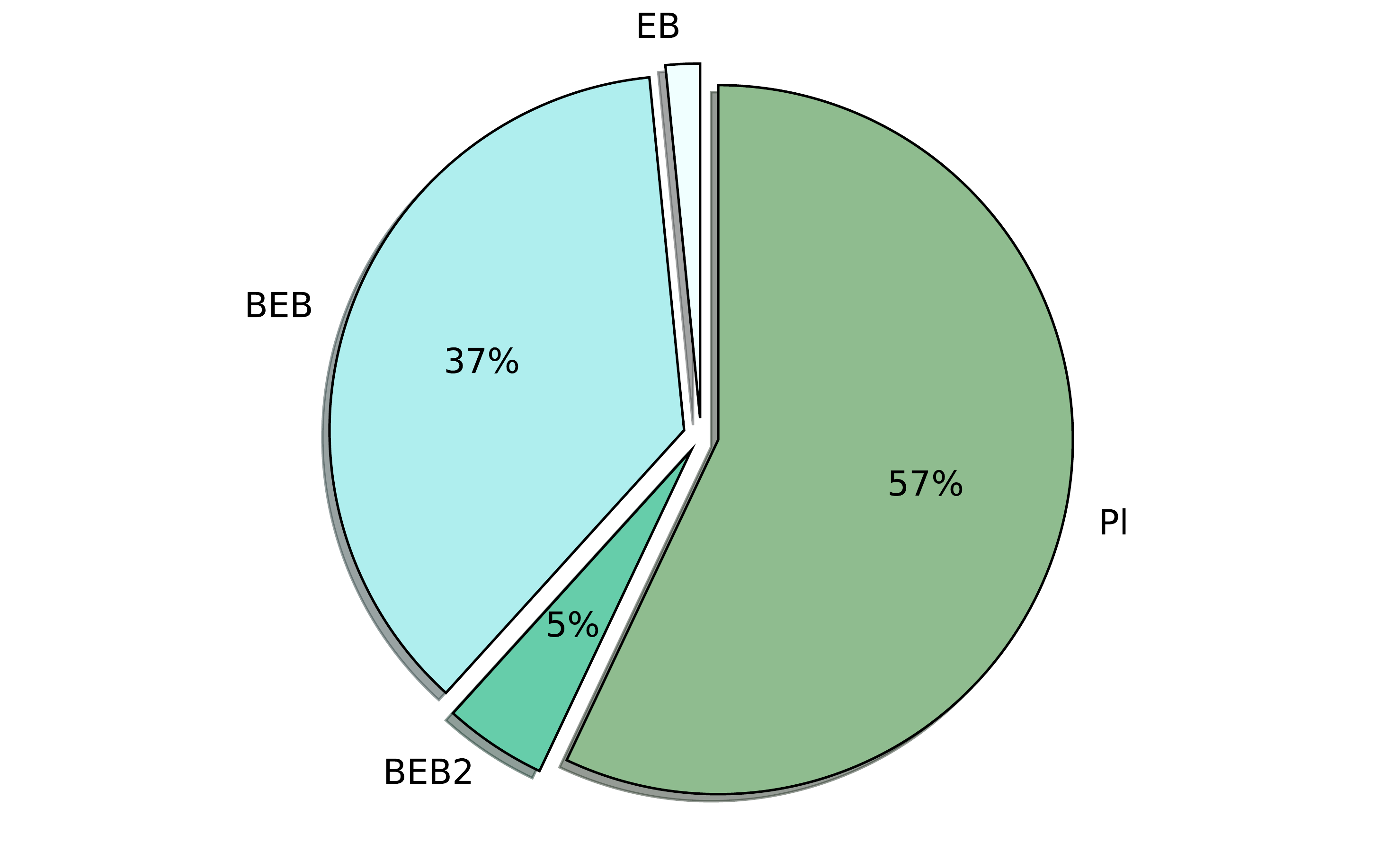}
    \caption{Most likely scenario for the 128 TOIs orbiting solar-analogue stars analysed. Each slice represents a different scenario simulated with \textsc{vespa} (Background Eclipsing Binary, BEB; Background Eclipsing Binary with period double, BEB2; Eclipsing Binary, EB; Planet, Pl) and shows the percentage of candidates for which such scenario is the most probable to occur. The EB slice corresponds to 2\%.}
    \label{fig:fpp-pie2}
\end{figure}

Among the entire selected sample, \textsc{vespa} was unable to evaluate some candidates due to problems related to the geometry of their orbital configuration or to difficulties in modelling their light curves (see Section \ref{sec:prob}). This happened in the most difficult cases, where the signal/noise ratio was very low. Therefore, the results we present here do not take them into account. Considering this removal plus the stars omitted in the previous section, we remain with 128 TOIs with usable light curves and reliable parameters. Among them, there are 23 candidates with a very high probability of being transiting planets, while almost 45 per cent of the entire sample have probability of being an FP that exceed 50 per cent (Fig. \ref{fig:fpp-pie}). Among the FPs, there are 26 candidates with FPP > 90\%. The remaining 48 candidates have an FPP with an intermediate value and their true transit nature requires further analysis to be confirmed.
The  histogram in Figure \ref{fig:isto} illustrates the FPP distribution of our candidates. In particular, it is possible to note that the FPP covers nearly the full 0-100 per cent probability range, with a higher concentration at the two extremes of the distribution.
Another important result concerns the FP scenario that appears to be the most recurrent one, i.e., not necessarily the one with a probability that exceed 50 per cent, but the one with the highest probability among all the scenarios analysed. As we can see from the pie chart in Figure \ref{fig:fpp-pie2}, the Background (or Foreground) Eclipsing Binary (BEB) is the main cause of FPs. This result is consistent with the expectations of the \textit{TESS} mission, for which the main cause of FPs is expected to be the BEB scenario \citep{2015JATIS...1a4003R}. This is caused by the crowding of stars within the \textit{TESS} photometric aperture, as a result of the large pixel size and the overall PSF area. In fact, having many light sources in the same photometric area may dilute the brightness of the observed source and increase the FP probability due to blended eclipsing binaries.

\subsection{Vetted candidates}
\label{valid}
\begin{figure*}
	\includegraphics[width=\textwidth]{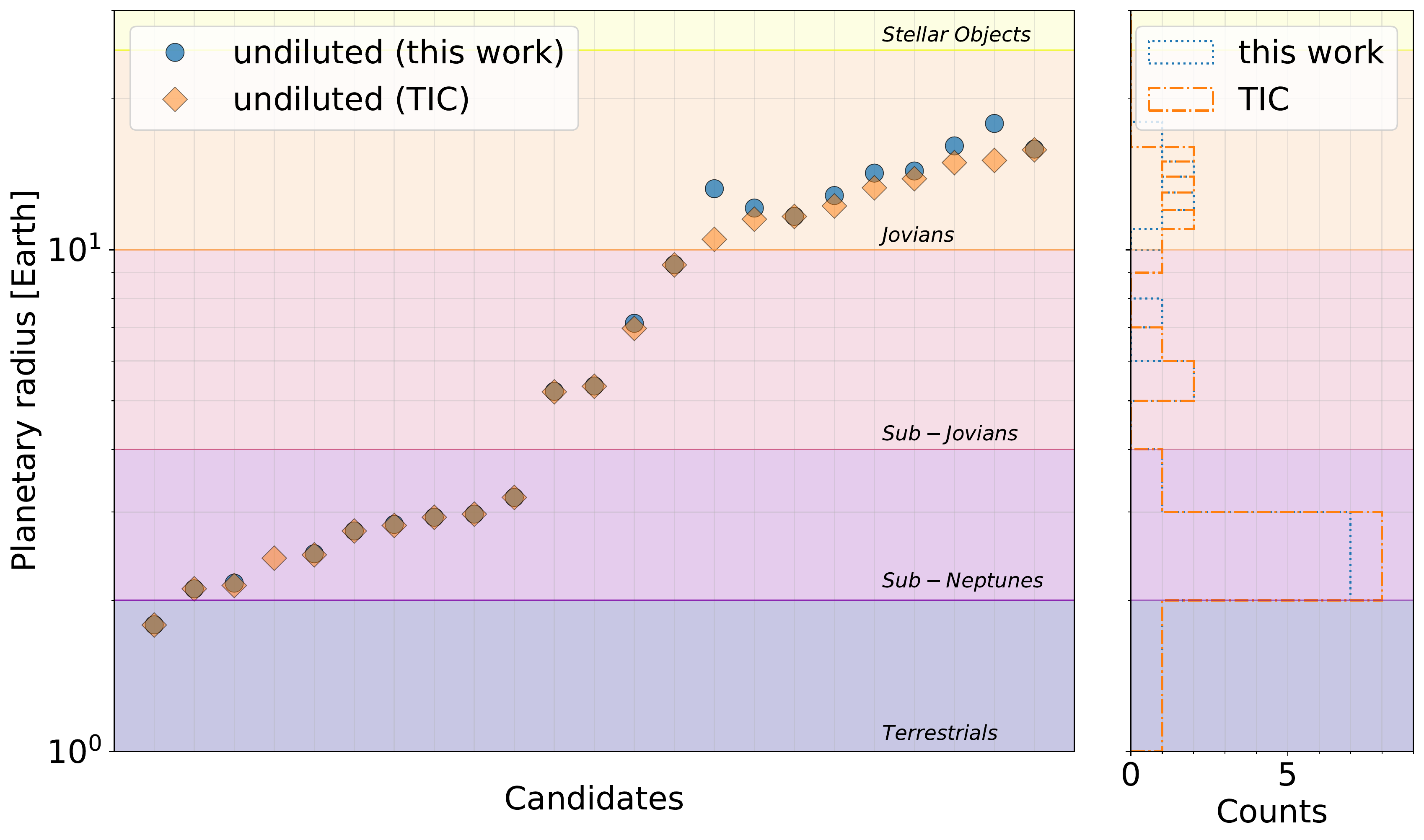}
    \caption{Planetary radius of our 23 statistically vetted candidates. On the x-axis, we represented a specific candidate, while on the y-axis its radius in units of Earth radii. We added five shaded areas to highlight different planet-size bins. The orange diamonds represent planetary radii from the \textit{TESS} mission, while the blue dots represent our evaluated parameters -- which are the radii corrected for the luminosity contamination coming from neighbourhood stars. Vertical lines are only guide for the eye. On the right panel, we plotted the planetary radii distribution. }
    \label{fig:radii}
\end{figure*}

To claim a statistical vetting for a transiting exoplanet candidate, we considered the FPP < 1\% threshold, as was done by \citet{2016ApJ...822...86M}. The number of candidates orbiting solar-analogue stars that satisfy this limit is 23, which corresponds to 18 per cent of TOIs within our selection (complete list in Table \ref{tab:validated}). Then, we subdivided the entire sample of statistically vetted candidates in five arbitrary planet-size bins: 
\begin{itemize}[leftmargin=*]
    \item Terrestrials: $R_p$ $\leqslant$ 2 $R_{\earth}$;
    \item Sub-Neptunes: 2$R_{\earth}$ < $R_p$ $\leqslant$ 4 $R_{\earth}$;
    \item Sub-Jovians: 4$R_{\earth}$ < $R_p$ $\leqslant$ 10 $R_{\earth}$;
    \item Jovians: 10$R_{\earth}$ < $R_p$ $\leqslant$ 25 $R_{\earth}$;
    \item Stellar objects: $R_p$ $\geqslant$ 25 $R_{\earth}$,
\end{itemize} to determine which kind of exoplanets we found. For each candidate, we considered two different values for their planetary radius. First of all, we took into account the radius estimated by the \textit{TESS} mission \citep{2019AJ....158..138S}. Then, we considered our estimated radius, which we obtained after performing a correction for the stellar dilution (see Sec. \ref{sec:undiluted}). In Figure \ref{fig:radii} we show the planetary radii of our sample of vetted candidates and its distribution. We added five shaded areas to highlight different planet-size bins. We can note the presence of one terrestrial-size exoplanet and the high concentration of Sub-Neptunes and Jovian-size candidates. At small radii, our estimated radii are quite similar to those coming from the \textit{TESS} mission, while they are often larger than the other estimate when the planetary radii are in the Jovian-size bin. The difference between our radii and the ones estimated by the \textit{TESS} mission might be due to the points highlighted in Sec. \ref{sec:undiluted} and may depend only on how the stellar dilution is treated (i.e., there is no dependence on the stellar radius). The two estimates are different (i.e., their difference is greater than $1\sigma$) for $\approx$ 17 per cent of the vetted candidates. We noticed that each of these vetted candidates is in the Jovians size bin and has been identified with the QLP pipeline. It is worth being aware of this difference. In fact, not only the planet-size bin could be different, but also the planetary nature of a candidate could become questionable if its radius reaches a specific value. We chose an arbitrary upper limit of 25 $R_{\earth}$ for a sub-stellar object, as the largest confirmed transiting exoplanet discovered so far has a similar size\footnote{Information from the \textit{NASA Exoplanet Archive}.} \citep{2017AJ....153..211Z}. We confirm that each of our vetted candidates has a sub-stellar radius.

\subsection{Vetted candidates confirmed to orbit their host star}
\label{sec:vett-conf}
Following the procedure described in Section \ref{stellar-neighbourhood}, we took advantage of \textit{Gaia} EDR3 photometry to accurately analyse the stellar neighbourhood of each vetted candidate. In this way, we were able to check whether any neighbourhood star could mimic the detected transit signal and successively exclude each \textit{Gaia} source as a possible BEB. This procedure allowed us to narrow the transit source origin of ten vetted candidates (see Section \ref{next} and Table \ref{tab:validated}) within 1.5 arcseconds separation from their host star. For the other 13 candidates, we have not been able to narrow down the location of the source of their signal. However, we have identified which neighbourhood stars might be a contaminant source and how deep their transit signal is. 

\subsubsection{Centroid motion results}
\begin{figure*}
	\includegraphics[width=\textwidth]{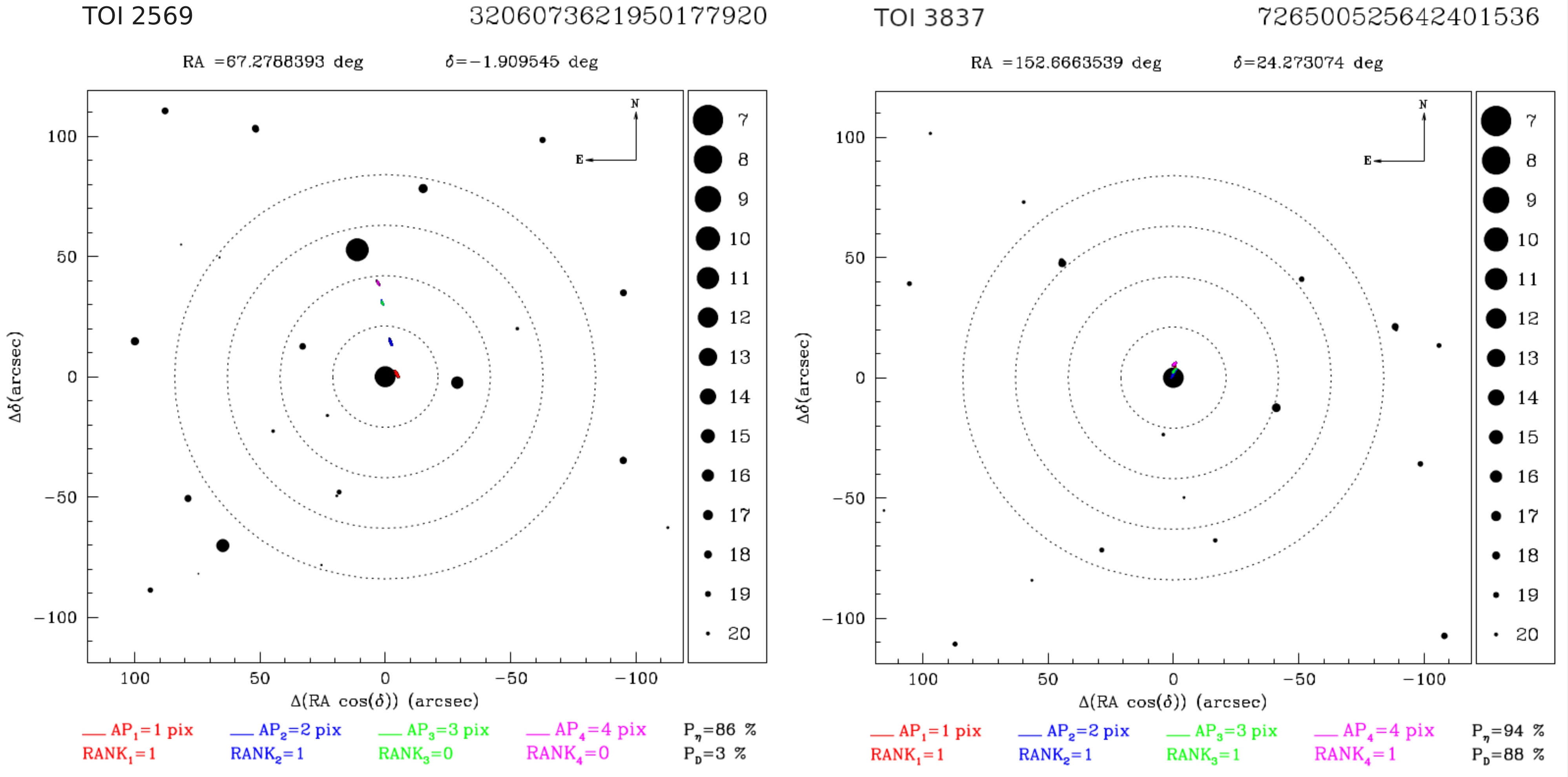}
    \caption{Two application examples of the centroid algorithm that is explained in \citet{2020MNRAS.498.1726M}. The left plot is centred on the star TOI 2569 and illustrates a failed centroid test. The colour-coded ellipses represent the position and dispersion of the centroid metric measurements relative to this target for four concentric apertures, as discussed in the paper we mentioned. We also showed the probability of source association $\rm P_{\rm D}$, which is equal to 3 per cent for this target. The right plot is centred on the star TOI 3837 and shows an example of a test passed. The probability of source association is 88 per cent for this target. } 
    \label{fig:centroid}
\end{figure*}

As additional evidence of the transit source origin of our vetted candidates, we considered the centroid motion test (see Section \ref{centroid}). In Table \ref{tab:validated} -- in the \textit{centroid test} column -- we show the results of this examination. Aside from a controversial case\footnote{The \textit{TESS} Data Validation reported a possible stellar contamination from a neighbour star.}, every vetted candidate whose on-target probability has been greatly enhanced -- with our \textit{Gaia} photometry analysis (see Section \ref{stellar-neighbourhood}) -- has passed the centroid motion test. Figure \ref{fig:centroid} presents an example of a test passed and one of a test failed. Summing the results of this analysis with those obtained using \textit{Gaia} photometry, we greatly enhanced the on-target probability of 12 vetted candidates.

\subsubsection{High-resolution imaging data}
To confirm the transit source origin of our vetted candidates, we consider the high-resolution Speckle/Adaptive Optics (AO) imaging data publicly available on the ExoFOP-TESS website (either as a table data or as an `Open Observing Note'). We need these follow-ups data to rule out unresolved neighbour stars beyond the 1.5 arcseconds spatial resolution of \textit{Gaia} EDR3. Summing the information gained with these data (further explained in Table \ref{tab:validated}) with those obtained using \textit{Gaia} photometry and centroid motion tests, we confirm the transit source origin of six vetted candidates and have greatly enhanced the on-target probability of another six vetted candidates. 

\subsection{On-off photometry observations}
\label{sec:on-off}
To perform the on-off photometry follow-up of our statistically vetted candidates (i.e., those from Sec. \ref{valid}), we have submitted an observational proposal to INAF AOT44 call (October 1st 2021 - March 31st, 2022, proposal REM-44018, P.I. Giacomo Mantovan), to collect multi-band REM images \citep{2003Msngr.113...40C, 2004AIPC..727..765M}. Located in La Silla, Chile, the REM telescope allows us to observe mainly the \textit{TESS} candidates detected in the southern hemisphere.

\begin{figure}
	\includegraphics[width=\columnwidth]{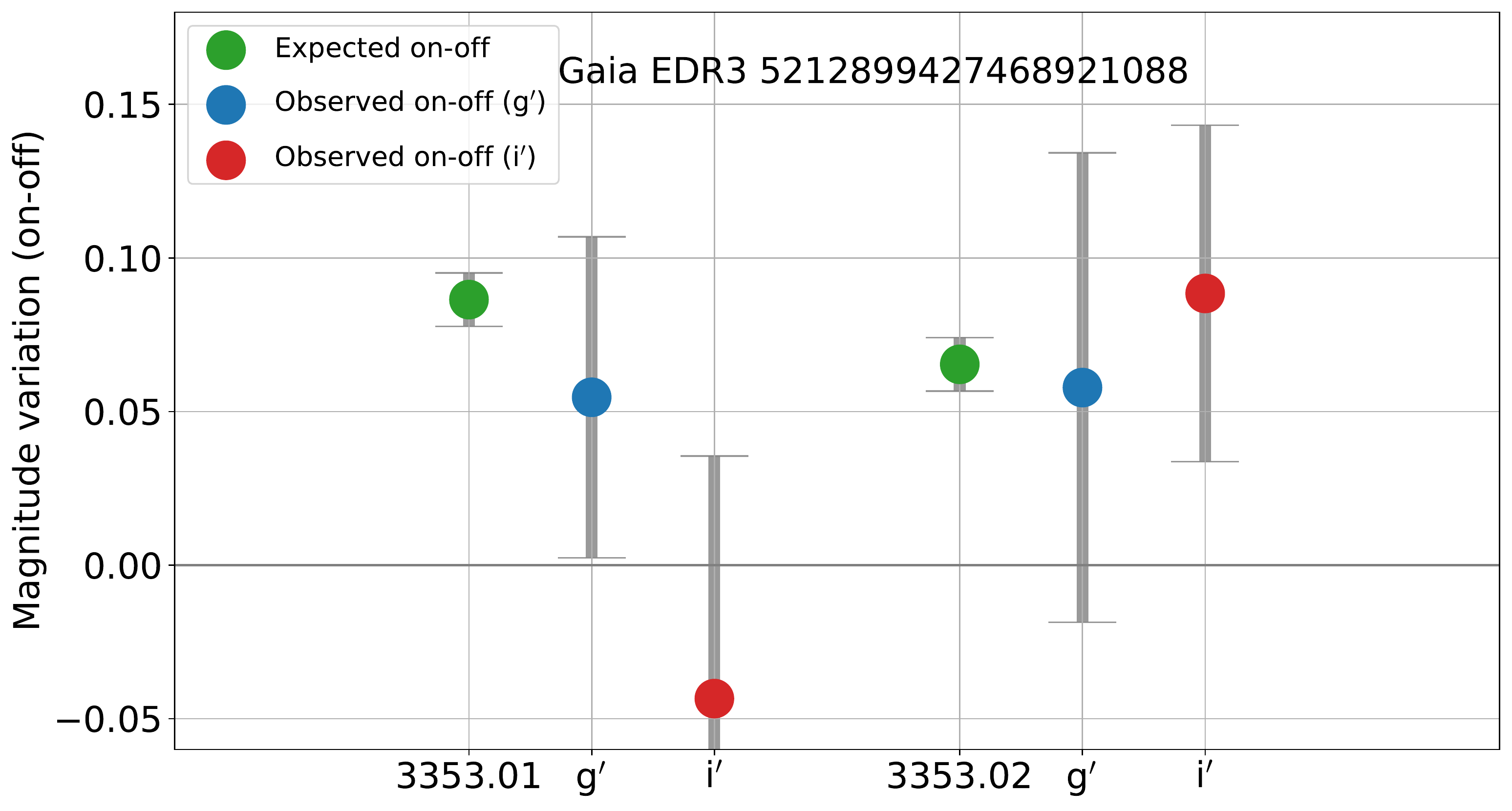}
    \caption{On-off magnitude variation of Gaia EDR3 5212899427468921088. The green dots represent the estimated on-off variation, while the blue and red ones represent the on-off observed magnitude variation in the Sloan/SDSS g$^\prime$ and i$^\prime$ filters, respectively. On the left-hand side, we show the magnitude variation during two predicted transit events of TOI 3353.01, which were observed with REM on the nights of 20 November 2021 and 4 December 2021. On the right-hand side, we show the magnitude variation during one expected transit event of TOI 3353.02, which was observed with REM on the night of 22 November 2021.}
    \label{fig:onoff}
\end{figure}

Thanks to these observations, we found that two candidates may be false positives. In particular, our analysis shows that both the transits of TOI 3353.01 and TOI 3353.02 could be due to a background eclipsing binary. In fact, Gaia EDR3 5212899427468921088  -- a neighbourhood star of TOI 3353 -- reproduces both the discovery signals (Fig. \ref{fig:onoff}). The magnitude variation has been calculated averaging the flux measured in several images during both the on- and off-transit phases, and correcting for systematic variations (i.e., different on- and off- zero point of magnitude due to different sky conditions occurring during the two phases). In addition, our procedure, which is a differential, aperture, transit photometry, removes most systematic trends, whereas we point out that some residual trends may still be present due to the individual, averaged, on-off measurement. We analysed REM images taken using the Sloan/SDSS g$^\prime$ filter and also the Sloan/SDSS i$^\prime$ filter, which allow us to perform a follow-up with an observation band similar to \textit{TESS} \citep{2015JATIS...1a4003R}. The observed -- and averaged -- on-off magnitude variation is comparable to the estimated one (eq. \ref{eq:magexp}). In addition to our analysis, the centroid motion tests carried out by the \textit{TESS} mission was unclear for both candidates, further suggesting a possible contamination. Moreover, the \textit{TESS} team -- in a note present on the ExoFOP \textit{TESS} website -- alerted the possible contamination for TOI 3353.02 exactly from the neighbour star we found.

Even though this result may seem reasonable and a lot of data suggests stellar contamination, we have reason to believe that this is a misleading result:
\begin{itemize}[leftmargin=*]
    \item the observed on-off variations present large error bars;
    \item we shortened both sides of the window length of TOI 3353.01's on-transit phase only by 1$\sigma$ (see Sec. \ref{stellar-neighbourhood}) because we were limited by the length of the data we had;
    \item when using REM images taken using the Sloan/SDSS i$^\prime$ filter, the observed on-off variation of the neighbourhood star does not reproduce the discovery signal of TOI 3353.01; 
    \item the target star is active and is also saturated in the REM photometry. These two aspects could affect the on-off photometry of the considered neighbourhood star, which is only 22 arcseconds far from the target; 
    \item there are no available data on the potential activity of Gaia EDR3 5212899427468921088. If this contaminating star were intrinsically variable, on-off photometry could give a `false negative';
    \item the orbital periods of TOI 3353.01 and TOI 3353.02 are not in phase with each other. We expect a 2:1 period commensurability if both transit were due to the same background eclipsing binary.
\end{itemize} Moreover, we independently reanalysed the \textit{TESS} light curves of TOI 3353.01 and TOI 3353.02, and modelled them using the \textsc{pycheops} code \citep{2021MNRAS.tmp.3057M}, to extrapolate the host star's stellar density $\rho_{*,h}$ from the transit signals. We followed equations 27 and 30 from \citet{2010arXiv1001.2010W}, and then ran MCMC simulations to better estimate the value and uncertainty of $\rho_{*,h}$. We compared $\rho_{*,h}$ with the nominal stellar density $\rho_{*}$, i.e., the one calculated from the stellar radius $R_*$ and mass $M_*$. Furthermore, we performed the same analysis focusing on the contaminant star, to determine if the resulting stellar density $\rho_{*,c}$ could better match the nominal value of the contaminant star. To do so, we injected the `third light' parameter (l\_3 in \textsc{pycheops}) into the \textsc{pycheops} modelling procedure, and treated the target star as the `third light' for the hypothetical transits in the contaminant. These analyses show that:
\begin{itemize}[leftmargin=*]    
    \item the resulting stellar density $\rho_{*,h}$ leads to values in agreement with the nominal stellar density $\rho_* = 0.94\pm0.18 \rho_{\sun}$ of the host star. In particular, we obtained $\rho_{*,h} = 0.69\pm0.23 \rho_{\sun}$ and $\rho_{*,h} = 0.91\pm0.36 \rho_{\sun}$, for TOI 3353.01 and TOI 3353.02, respectively;
    \item the resulting stellar density $\rho_{*,c}$ leads to values larger than $\rho_{\sun}$, which are not consistent with the nominal stellar density $\rho_* = 0.24 \rho_{\sun}$ of the contaminant star. Given the large magnitude difference between the host and the contaminant star, treating the target as the `third light' for the hypothetical transits in the contaminant leads the model to produce transits so deep that the occulting body has to be very large. This is contradicted by the observations, where the short duration of ingress and egress mandates a much smaller body.
\end{itemize}

These considerations imply that the results of our on-off analysis are not accurate enough to confirm the source of transit signals or detect the source of contamination. We can also rule out the physical scenario of a contaminating eclipsing binary capable of explaining both transiting candidates. Moreover, through the analysis of stellar density from both transit models, we have reasons to believe that TOI 3353 is a genuine multi-planetary system. Regardless of the latter result, we emphasise that further photometric observations are crucial to shed light on the true nature of these two candidates. In particular, we suggest performing full-transit photometric observations and focusing attention on the neighbourhood star Gaia EDR3 5212899427468921088.



\subsection{Problematic cases}
\label{sec:prob}
The \textsc{vespa} code was unable to evaluate 21 candidates orbiting solar-analogue stars within our sample. In more detail:
\begin{enumerate}[leftmargin=*]
    \item Three candidates have orbital period and simulated stellar properties that imply their orbit to be within their host star's Roche limit. \textsc{vespa} considers this situation as a FP;
    \item Eight TOIs did receive MCMC modelling but the trapezoidal fitting model was not able to fit the transit signal. All these candidates have a very low signal/noise ratio;
    \item The trapezoid MCMC fit did not converge for one candidate;
    \item Nine TOIs have no light-curve publicly-available on the MAST portal.
\end{enumerate}

\section{Discussion}
\label{sec:discussion}
The validation process is a fundamental part of a more complex work, which aims to identify false positives and exclude them to obtain a cleaner sample of candidates, leading to the final confirmation of an exoplanet candidate. Thanks to this procedure, we can eliminate most of the false candidates, and proceed with the follow-up observations up to the radial velocity measurement, which allows the full characterisation of a planetary system. Only after the radial velocity procedure, it is possible to confirm an exoplanet candidate. 

\subsection{Follow-up observations}
\label{next}
In this work, we statistically vetted 23 \textit{TESS} candidates orbiting solar-analogue stars and subsequently analysed their stellar neighbourhood to investigate the presence of possible contaminant stars, confirming the transit source origin of six of them and greatly enhancing the on-target probability of another six of them. These two steps allow us to determine which are the best targets and which are the next follow-up observations required to fully confirm and characterise their planetary nature. 

For some of our targets, the high-resolution spectra of their host stars are available in public archives, and these are listed in detail in Table \ref{tab:spectra}. If the name of the host star is duplicated, it shows that there are more than one instrument that has obtained the spectrum of the host star. All these instruments have been included in Table \ref{tab:spectra}. For some of our targets, there are RV measurements performed with low-resolution spectroscopy. However, we do not report on the spectra obtained with those instruments but only the `Open Observing Notes' on these RV measurements that are publicly available on the ExoFOP website. It should be noted that, however, there are no high-precision RV measurements reported for any of these stars, which would lead to confirming or ruling out an exoplanet candidate.

Depending on which instrument has published the spectra, and with what spectral coverage and resolution, the next steps for follow-up on our targets will be determined. For our targets, as incorporated and specified in Table \ref{tab:spectra}, the available spectra are obtained with TRES \citep{2007RMxAC..28..129S} (resolution $\sim$ 44000), FIDEOS \citep{2014SPIE.9147E..89T} (resolution $\sim$ 43000), and CHIRON \citep{2013PASP..125.1336T} (resolution $\sim$ 80000). Based on the capabilities of each instrument, different follow-up paths should be pursued for each of our targets, as we will detail later in this Section. It should be noted, however, that RV confirmation requires resource-intensive long-term monitoring programs. On ExoFOP catalogue, all the spectra obtained by CHIRON (which have the highest resolution in Table \ref{tab:spectra}) for our targets are flagged as not appropriate to precision RV (PRV), which is necessary for directly measure the stellar reflex motion due to planets and derive planet masses. We hence focus on RV follow-up strategies that have not been yet conducted for our targets. 

Low-precision RV measurements are necessary to reject grazing eclipsing binaries or transiting massive white and brown dwarfs, which can not be identified through the transit method. It is also essential because often the presence of a stellar body can be ruled out after taking three radial velocity observations. This aim can be fulfilled by either TRES or FIDEOS, as listed in Table \ref{tab:spectra}. In fact, TRES and FIDEOS are usually used for identifying the nature of the transiting objects or ruling out false positives (a technique known as \textit{reconnaissance spectroscopy}). Any candidate that will survive this test will become an exquisite target for internal structure and atmospheric characterisation through high-precision RV measurements, which can be conducted by higher resolution spectrometers -- that confirm exoplanet candidates by determining their masses -- such as CHIRON. 

\subsubsection{Definition of `statistically validated planets' and priority marks}
\label{sec:definition}
Considering all the above points and state-of-the-art planet validation papers, we recognise some TOIs investigated in this work as fully `statistically validated planets'. For such TOIs we have added a small letter planet suffix (e.g., `b', `c') which takes the place of .01/.02 previously present (see Table \ref{tab:validated}). Any other TOI investigated in this work is instead referred to as `vetted'. In particular, only TOIs meeting the following criteria should be given a planet suffix:
\begin{itemize}[leftmargin=*]
    \item The transit signal has been confirmed to be on-target (i.e., relative to a \textit{maxrad} area that contains no known neighbouring stars bright enough to cause the event, including Adaptive Optics/Speckle neighbours);
    \item Host star spectroscopy should not be suggestive of a composite spectrum, a large RV offset indicative of an EB, or an RV orbit that is out-of-phase with the photometric ephemeris;
    \item The TOI should have a well sampled transit shape (i.e., high photometric precision, high number of transits observed, short cadence sampling or transits very deep) to be used for the statistical validation;
    \item The TOI has been probabilistically vetted with FPP < 0.01;
    \item The TOI has a uniquely determined orbital period.
\end{itemize}

Moreover, here we suggest a priority mark to establish what the next step is for the five statistically validated planets and the 18 exoplanet candidates vetted in this paper (see Table \ref{tab:validated}):
\begin{itemize}[leftmargin=*]
\item Mark = 1: candidate with greatly enhanced on-target probability, and both low-precision RV measurements and high-resolution imaging data are already available. High-precision RV observations should be conducted; 
\item Mark = 2: candidate with greatly enhanced on-target probability, and low-precision RV measurements are already available. High-resolution Adaptive Optics/Speckle imaging observations should be conducted;
\item Mark = 3: candidate with greatly enhanced on-target probability, but no (or not enough) RV measurements are available. We suggest performing low-precision RV observations;
\item Mark = 4: candidate whose on-target probability is low or unclear. We suggest to perform on-off photometry observations. 
\end{itemize} 

\begin{figure}
	\includegraphics[width=\columnwidth]{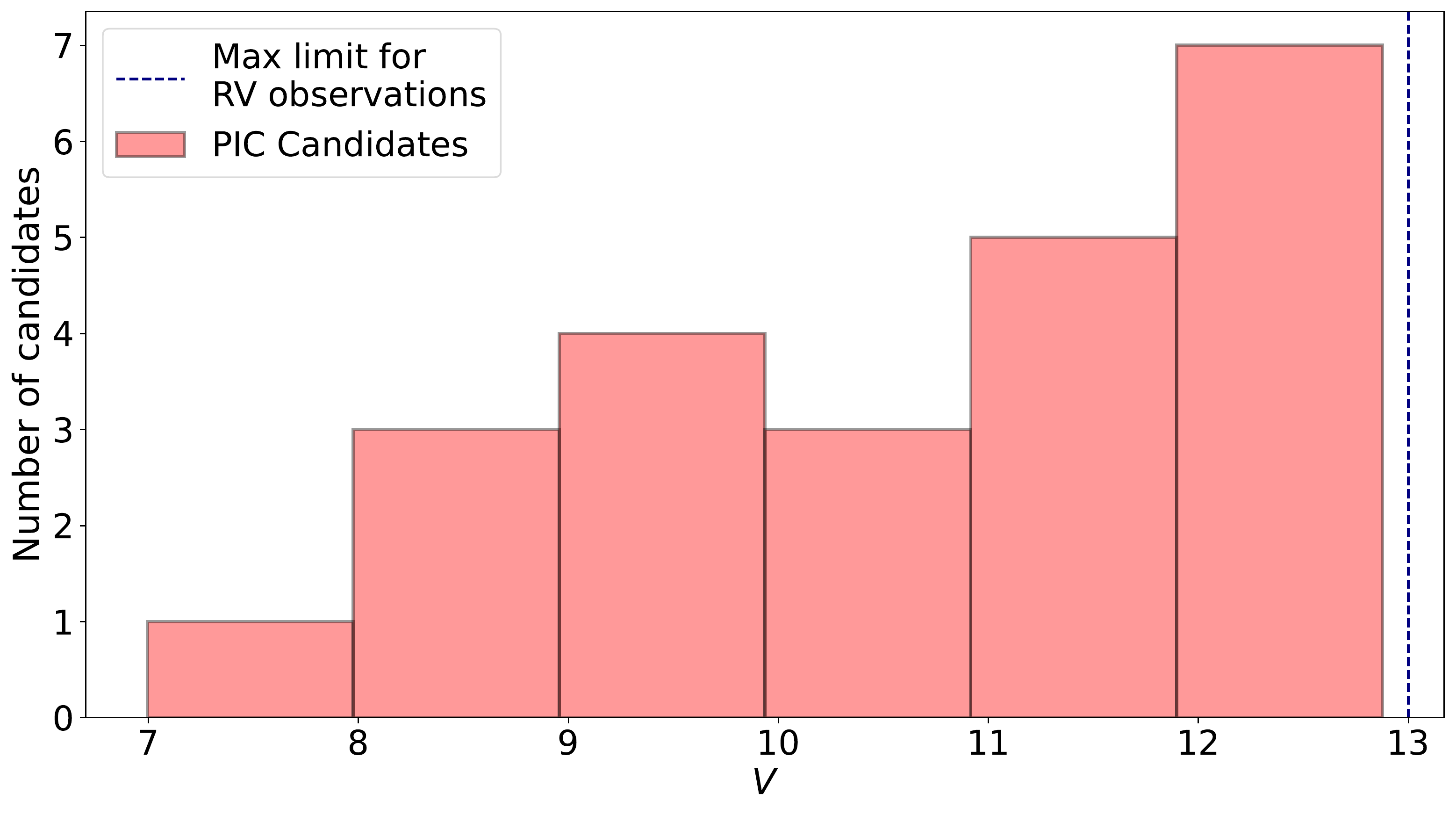}
    \caption{Distribution in V magnitude of our 23 vetted candidates.}
    \label{fig:V}
\end{figure}

\subsection{Statistical validation reliability}
We acknowledge that our statistical validation analysis strongly relies on the stellar neighbourhood analysis. In our calculation, we have added a constraint (\textit{maxrad}) to account for the probability that the transit originates from the target. However, the use of \textit{Gaia} EDR3 photometry, centroid motion tests or subsequent on-off analyses is necessary to give 100 per cent reliability to our results. Moreover, high-resolution imaging follow-ups are needed to rule out unresolved neighbour stars beyond the 1.5 arcseconds spatial resolution of \textit{Gaia} EDR3. In fact, if all neighbour stars are not adequately ruled out as transit sources prior to the analysis, \textsc{vespa} could classify false positive candidates as true planets \citep{2020arXiv200200691G, 2016ApJ...822...86M}. For this reason, we performed the \textit{Gaia} EDR3 photometry analysis and the centroid motion tests before starting the \textsc{vespa} code, and we checked the high-resolution imaging data publicly available on the ExoFOP website. We require subsequent on-off analyses for those candidates not yet confirmed orbiting their host star. This emphasises the importance of careful consideration of potential contamination from the host star’s neighbourhood (see, for example, the notes in Table \ref{tab:validated} on the on-target probability of TOI 1689.01), then precisely following the priority marks we have suggested (Sect. \ref{next}) to fully validate and later confirm our target exoplanets. We emphasize that our candidates with Mark = 1 are fully validated planets and ready to be confirmed.

In contrast, the \textit{maxrad} constraint that we have adopted enhances -- by construction -- the probability of the BEB scenario, and hence the total FPP of a candidate (also noted by \citealt{2021MNRAS.508..195D}). Therefore, we can only identify the planetary candidate after excluding the BEB scenario through a complete analysis of the stellar neighbourhood. This is for example the case of TOI 4399.01, which was first identified by \textsc{vespa} as a possible BEB and then as a likely planet after ruling out contaminant stars using \textit{Gaia} EDR3 photometry.

\section{Conclusions}
\label{sec:conclusion}
Here we presented our ongoing follow-up program of \textit{TESS} candidates orbiting solar-analogue stars that are in the all-sky PLATO input catalogue. Our probabilistic validation analysis allows us to identify which are the most promising candidates, while the evaluation of their stellar neighbourhood determines which are the next follow-up observations needed to confirm their exoplanet nature. The final goal of the entire procedure is to avoid wasting observational time at expensive facilities and optimize follow-up resources. In particular, we statistically vetted 23 \textit{TESS} candidates orbiting solar-analogue stars. Five of them have been confirmed on-target and are ready for follow-up high-precision radial velocity observations (we refer to them as `statistically validated planets'), another three have a greatly enhanced on-target probability and need high-resolution imaging data, while the others need additional spectroscopic and/or photometric observations (see Table \ref{tab:validated}, column `Priority \& obs.'). 

It is worth noting that these are new discoveries. We will continue to search for new validated planets at least as long as the \textit{TESS} mission will continue. In the very near future, we will complete the on-off photometry follow-up of our best targets, by proposing further investigation with REM and other telescopes, as well as low-precision radial velocity observations. This will allow us to extend the sample of vetted candidates and hence the number of genuine targets to be later characterized through high-precision radial velocity observations. Similarly to \textit{TESS}, the future PLATO transit mission \citep{2014ExA....38..249R} will have a low spatial resolution (15 arcsec/pixel,  \citealt{2017SPIE10564E..05L}); hence, it will also require a quick and efficient statistical validation procedure to exclude false positives from the large number of candidates that PLATO will discover. To conclude, our validation procedure will be essential and should be rather easily adaptable to the future PLATO mission.\newline

The authors became aware of the confirmation of TOI 4399 b \citep{2022AJ....163..289Z} during the referee process. This independent work verifies our process by confirming one of our five validated planets. Moreover, we want to bring attention to the follow-up work of TOI-5398 on TRES by Jiayin Dong et al. (private communication), which allowed the establishment of a tentative orbit. This independent work further demonstrates that our method finds good targets for follow-up.

\section*{Acknowledgements}
We are extremely grateful to the anonymous referee for the thorough comments, which undoubtedly improved the quality of this manuscript. We acknowledge Dr. David Latham and Allyson Bieryla for providing helpful in-depth interpretation of TRES observations. We also acknowledge Dr. Boris Safonov for the precious discussion on TOI 1689. I would like to acknowledge the contribution of Filippo Santoliquido in helping me to improve the scientific impact of many figures. I would also like to acknowledge the contribution of Ho-Hin Leung, who helped me improving the readability of the text.

G.M. acknowledges the support of the Erasmus+ Programme of the European Union and of the doctoral grant funded by the University of Padova and by the Italian Ministry of Education, University and Research (MIUR). G.M. is also grateful to the Centre for Exoplanet Science, University of St Andrews (StA-CES) for hospitality and computing resources. GPi, LBo, VNa, and FZM acknowledge the funding support from Italian Space Agency (ASI) regulated by `Accordo ASI-INAF n. 2013-016-R.0 del 9 luglio 2013 e integrazione del 9 luglio 2015 CHEOPS Fasi A/B/C'. We acknowledge the support of PLATO ASI-INAF agreements n.2015-019-R0-2015 and n. 2015-019-R.1-2018. T.G.W. and A.C.C. acknowledge support from STFC consolidated grant number ST/V000861/1, and UKSA grant ST/R003203/1. This research has made use of the Exoplanet Follow-up Observation Program (ExoFOP; DOI: 10.26134/ExoFOP5) website, which is operated by the California Institute of Technology, under contract with the National Aeronautics and Space Administration under the Exoplanet Exploration Program. We acknowledge the use of TESS High Level Science Products (HLSP) produced by the Quick-Look Pipeline (QLP) at the TESS Science Office at MIT, which are publicly available from the Mikulski Archive for Space Telescopes (MAST). Funding for the TESS mission is provided by NASA's Science Mission directorate. 

\textit{Software}: \textsc{Astropy} \citep{2013A&A...558A..33A}, \textsc{astroquery} \citep{2019AJ....157...98G}, \textsc{emcee} \citep{2013PASP..125..306F}, \textsc{Isochrones} \citep{2015ascl.soft03010M}, \textsc{lightkurve} \citep{2018ascl.soft12013L}, \textsc{matplotlib} \citep{2018ascl.soft12013L}, \textsc{Numpy} \citep{2011CSE....13b..22V}, \textsc{pandas} \citep{jeff-reback-2021-5501881}, \textsc{pycheops} \citep{2021MNRAS.tmp.3057M}, \textsc{MultiNest} \citep{2019OJAp....2E..10F},  \textsc{Scipy} \citep{2007CSE.....9c..10O}, \textsc{vespa} \citep{2012ApJ...761....6M}. 

This work is dedicated to my beloved grandma.

\section*{Data Availability}

The authors confirm that the data supporting the findings of this study are available within the article and its supplementary materials. 
 



\bibliographystyle{mnras}
\bibliography{references} 

\begin{thebibliography}{}
\makeatletter
\relax
\def\mn@urlcharsother{\let\do\@makeother \do\$\do\&\do\#\do\^\do\_\do\%\do\~}
\def\mn@doi{\begingroup\mn@urlcharsother \@ifnextchar [ {\mn@doi@}
  {\mn@doi@[]}}
\def\mn@doi@[#1]#2{\def\@tempa{#1}\ifx\@tempa\@empty \href
  {http://dx.doi.org/#2} {doi:#2}\else \href {http://dx.doi.org/#2} {#1}\fi
  \endgroup}
\def\mn@eprint#1#2{\mn@eprint@#1:#2::\@nil}
\def\mn@eprint@arXiv#1{\href {http://arxiv.org/abs/#1} {{\tt arXiv:#1}}}
\def\mn@eprint@dblp#1{\href {http://dblp.uni-trier.de/rec/bibtex/#1.xml}
  {dblp:#1}}
\def\mn@eprint@#1:#2:#3:#4\@nil{\def\@tempa {#1}\def\@tempb {#2}\def\@tempc
  {#3}\ifx \@tempc \@empty \let \@tempc \@tempb \let \@tempb \@tempa \fi \ifx
  \@tempb \@empty \def\@tempb {arXiv}\fi \@ifundefined
  {mn@eprint@\@tempb}{\@tempb:\@tempc}{\expandafter \expandafter \csname
  mn@eprint@\@tempb\endcsname \expandafter{\@tempc}}}

\bibitem[\protect\citeauthoryear{{Astropy Collaboration} et~al.,}{{Astropy
  Collaboration} et~al.}{2013}]{2013A&A...558A..33A}
{Astropy Collaboration} et~al., 2013, \mn@doi [\aap]
  {10.1051/0004-6361/201322068}, \href
  {https://ui.adsabs.harvard.edu/abs/2013A&A...558A..33A} {558, A33}

\bibitem[\protect\citeauthoryear{Bradley et~al.,}{Bradley
  et~al.}{2021}]{larry-bradley-2021-5525286}
Bradley L.,  et~al., 2021, astropy/photutils: 1.2.0,
  \mn@doi{10.5281/zenodo.5525286}, \url
  {https://doi.org/10.5281/zenodo.5525286}

\bibitem[\protect\citeauthoryear{{Caldwell} et~al.,}{{Caldwell}
  et~al.}{2020}]{2020RNAAS...4..201C}
{Caldwell} D.~A.,  et~al., 2020, \mn@doi [Research Notes of the American
  Astronomical Society] {10.3847/2515-5172/abc9b3}, \href
  {https://ui.adsabs.harvard.edu/abs/2020RNAAS...4..201C} {4, 201}

\bibitem[\protect\citeauthoryear{{Chincarini} et~al.,}{{Chincarini}
  et~al.}{2003}]{2003Msngr.113...40C}
{Chincarini} G.,  et~al., 2003, The Messenger, \href
  {https://ui.adsabs.harvard.edu/abs/2003Msngr.113...40C} {113, 40}

\bibitem[\protect\citeauthoryear{{Choi}, {Dotter}, {Conroy}, {Cantiello},
  {Paxton}  \& {Johnson}}{{Choi} et~al.}{2016}]{2016ApJ...823..102C}
{Choi} J.,  {Dotter} A.,  {Conroy} C.,  {Cantiello} M.,  {Paxton} B.,
  {Johnson} B.~D.,  2016, \mn@doi [\apj] {10.3847/0004-637X/823/2/102}, \href
  {https://ui.adsabs.harvard.edu/abs/2016ApJ...823..102C} {823, 102}

\bibitem[\protect\citeauthoryear{{De Leon} et~al.,}{{De Leon}
  et~al.}{2021}]{2021MNRAS.508..195D}
{De Leon} J.~P.,  et~al., 2021, \mn@doi [\mnras] {10.1093/mnras/stab2305},
  \href {https://ui.adsabs.harvard.edu/abs/2021MNRAS.508..195D} {508, 195}

\bibitem[\protect\citeauthoryear{{Deeg} et~al.,}{{Deeg}
  et~al.}{2009}]{2009A&A...506..343D}
{Deeg} H.~J.,  et~al., 2009, \mn@doi [\aap] {10.1051/0004-6361/200912011},
  \href {https://ui.adsabs.harvard.edu/abs/2009A&A...506..343D} {506, 343}

\bibitem[\protect\citeauthoryear{{Dotter}}{{Dotter}}{2016}]{2016ApJS..222....8D}
{Dotter} A.,  2016, \mn@doi [\apjs] {10.3847/0067-0049/222/1/8}, \href
  {https://ui.adsabs.harvard.edu/abs/2016ApJS..222....8D} {222, 8}

\bibitem[\protect\citeauthoryear{{Feroz}, {Hobson}, {Cameron}  \&
  {Pettitt}}{{Feroz} et~al.}{2019}]{2019OJAp....2E..10F}
{Feroz} F.,  {Hobson} M.~P.,  {Cameron} E.,   {Pettitt} A.~N.,  2019, \mn@doi
  [The Open Journal of Astrophysics] {10.21105/astro.1306.2144}, \href
  {https://ui.adsabs.harvard.edu/abs/2019OJAp....2E..10F} {2, 10}

\bibitem[\protect\citeauthoryear{{Foreman-Mackey}, {Hogg}, {Lang}  \&
  {Goodman}}{{Foreman-Mackey} et~al.}{2013}]{2013PASP..125..306F}
{Foreman-Mackey} D.,  {Hogg} D.~W.,  {Lang} D.,   {Goodman} J.,  2013, \mn@doi
  [\pasp] {10.1086/670067}, \href
  {https://ui.adsabs.harvard.edu/abs/2013PASP..125..306F} {125, 306}

\bibitem[\protect\citeauthoryear{{Gaia Collaboration} et~al.,}{{Gaia
  Collaboration} et~al.}{2021}]{2021A&A...649A...1G}
{Gaia Collaboration} et~al., 2021, \mn@doi [\aap]
  {10.1051/0004-6361/202039657}, \href
  {https://ui.adsabs.harvard.edu/abs/2021A&A...649A...1G} {649, A1}

\bibitem[\protect\citeauthoryear{{Giacalone} et~al.,}{{Giacalone}
  et~al.}{2020}]{2020arXiv200200691G}
{Giacalone} S.,  et~al., 2020, arXiv e-prints, \href
  {https://ui.adsabs.harvard.edu/abs/2020arXiv200200691G} {p. arXiv:2002.00691}

\bibitem[\protect\citeauthoryear{{Ginsburg} et~al.,}{{Ginsburg}
  et~al.}{2019}]{2019AJ....157...98G}
{Ginsburg} A.,  et~al., 2019, \mn@doi [\aj] {10.3847/1538-3881/aafc33}, \href
  {https://ui.adsabs.harvard.edu/abs/2019AJ....157...98G} {157, 98}

\bibitem[\protect\citeauthoryear{{Girardi}, {Groenewegen}, {Hatziminaoglou}  \&
  {da Costa}}{{Girardi} et~al.}{2005}]{2005A&A...436..895G}
{Girardi} L.,  {Groenewegen} M.~A.~T.,  {Hatziminaoglou} E.,   {da Costa} L.,
  2005, \mn@doi [\aap] {10.1051/0004-6361:20042352}, \href
  {https://ui.adsabs.harvard.edu/abs/2005A&A...436..895G} {436, 895}

\bibitem[\protect\citeauthoryear{{Guerrero} et~al.,}{{Guerrero}
  et~al.}{2021}]{2021arXiv210312538G}
{Guerrero} N.~M.,  et~al., 2021, arXiv e-prints, \href
  {https://ui.adsabs.harvard.edu/abs/2021arXiv210312538G} {p. arXiv:2103.12538}

\bibitem[\protect\citeauthoryear{{Huang} et~al.,}{{Huang}
  et~al.}{2020}]{2020RNAAS...4..204H}
{Huang} C.~X.,  et~al., 2020, \mn@doi [Research Notes of the American
  Astronomical Society] {10.3847/2515-5172/abca2e}, \href
  {https://ui.adsabs.harvard.edu/abs/2020RNAAS...4..204H} {4, 204}

\bibitem[\protect\citeauthoryear{{Jenkins} et~al.,}{{Jenkins}
  et~al.}{2016}]{2016SPIE.9913E..3EJ}
{Jenkins} J.~M.,  et~al., 2016, in {Chiozzi} G.,  {Guzman} J.~C.,  eds,
  Society of Photo-Optical Instrumentation Engineers (SPIE) Conference Series
  Vol. 9913, Software and Cyberinfrastructure for Astronomy IV. p. 99133E,
  \mn@doi{10.1117/12.2233418}

\bibitem[\protect\citeauthoryear{{Kane}, {Ciardi}, {Gelino}  \& {von
  Braun}}{{Kane} et~al.}{2012}]{2012MNRAS.425..757K}
{Kane} S.~R.,  {Ciardi} D.~R.,  {Gelino} D.~M.,   {von Braun} K.,  2012,
  \mn@doi [\mnras] {10.1111/j.1365-2966.2012.21627.x}, \href
  {https://ui.adsabs.harvard.edu/abs/2012MNRAS.425..757K} {425, 757}

\bibitem[\protect\citeauthoryear{Kass \& Raftery}{Kass \&
  Raftery}{1995}]{kass1995bayes}
Kass R.~E.,  Raftery A.~E.,  1995, Journal of the american statistical
  association, 90, 773

\bibitem[\protect\citeauthoryear{{Kipping}}{{Kipping}}{2013}]{2013MNRAS.434L..51K}
{Kipping} D.~M.,  2013, \mn@doi [\mnras] {10.1093/mnrasl/slt075}, \href
  {https://ui.adsabs.harvard.edu/abs/2013MNRAS.434L..51K} {434, L51}

\bibitem[\protect\citeauthoryear{Knuth, Habeck, Malakar, Mubeen  \&
  Placek}{Knuth et~al.}{2015}]{KNUTH201550}
Knuth K.~H.,  Habeck M.,  Malakar N.~K.,  Mubeen A.~M.,   Placek B.,  2015,
  \mn@doi [Digital Signal Processing]
  {https://doi.org/10.1016/j.dsp.2015.06.012}, 47, 50

\bibitem[\protect\citeauthoryear{{Kreidberg}}{{Kreidberg}}{2015}]{2015PASP..127.1161K}
{Kreidberg} L.,  2015, \mn@doi [\pasp] {10.1086/683602}, \href
  {https://ui.adsabs.harvard.edu/abs/2015PASP..127.1161K} {127, 1161}

\bibitem[\protect\citeauthoryear{{Laubier}, {Bodin}, {Pasquier}, {Fredon},
  {Levacher}, {Vola}, {Buey}  \& {Bernardi}}{{Laubier}
  et~al.}{2017}]{2017SPIE10564E..05L}
{Laubier} D.,  {Bodin} P.,  {Pasquier} H.,  {Fredon} S.,  {Levacher} P.,
  {Vola} P.,  {Buey} T.,   {Bernardi} P.,  2017, in Society of Photo-Optical
  Instrumentation Engineers (SPIE) Conference Series. p. 1056405,
  \mn@doi{10.1117/12.2309075}

\bibitem[\protect\citeauthoryear{{Lightkurve Collaboration}
  et~al.,}{{Lightkurve Collaboration} et~al.}{2018}]{2018ascl.soft12013L}
{Lightkurve Collaboration} et~al., 2018, {Lightkurve: Kepler and TESS time
  series analysis in Python}, Astrophysics Source Code Library, record
  ascl:1812.013 (\mn@eprint {ascl} {1812.013})

\bibitem[\protect\citeauthoryear{Mahalanobis}{Mahalanobis}{1936}]{mahalanobis1936generalized}
Mahalanobis P.~C.,  1936.

\bibitem[\protect\citeauthoryear{{Maxted} et~al.,}{{Maxted}
  et~al.}{2021}]{2021MNRAS.tmp.3057M}
{Maxted} P.~F.~L.,  et~al., 2021, \mn@doi [\mnras] {10.1093/mnras/stab3371},
  \href {https://ui.adsabs.harvard.edu/abs/2021MNRAS.tmp.3057M} {}

\bibitem[\protect\citeauthoryear{{Mayor} et~al.,}{{Mayor}
  et~al.}{2011}]{2011arXiv1109.2497M}
{Mayor} M.,  et~al., 2011, arXiv e-prints, \href
  {https://ui.adsabs.harvard.edu/abs/2011arXiv1109.2497M} {p. arXiv:1109.2497}

\bibitem[\protect\citeauthoryear{{Molinari}, {Vergani}, {Zerbi}, {Covino}  \&
  {Chincarini}}{{Molinari} et~al.}{2004}]{2004AIPC..727..765M}
{Molinari} E.,  {Vergani} S.~D.,  {Zerbi} F.~M.,  {Covino} S.,   {Chincarini}
  G.,  2004, in {Fenimore} E.,  {Galassi} M.,  eds,  American Institute of
  Physics Conference Series Vol. 727, Gamma-Ray Bursts: 30 Years of Discovery.
  pp 765--768, \mn@doi{10.1063/1.1810954}

\bibitem[\protect\citeauthoryear{{Montalto} et~al.,}{{Montalto}
  et~al.}{2020}]{2020MNRAS.498.1726M}
{Montalto} M.,  et~al., 2020, \mn@doi [\mnras] {10.1093/mnras/staa2438}, \href
  {https://ui.adsabs.harvard.edu/abs/2020MNRAS.498.1726M} {498, 1726}

\bibitem[\protect\citeauthoryear{{Montalto} et~al.,}{{Montalto}
  et~al.}{2021}]{2021A&A...653A..98M}
{Montalto} M.,  et~al., 2021, \mn@doi [\aap] {10.1051/0004-6361/202140717},
  \href {https://ui.adsabs.harvard.edu/abs/2021A&A...653A..98M} {653, A98}

\bibitem[\protect\citeauthoryear{{Morton}}{{Morton}}{2012}]{2012ApJ...761....6M}
{Morton} T.~D.,  2012, \mn@doi [\apj] {10.1088/0004-637X/761/1/6}, \href
  {https://ui.adsabs.harvard.edu/abs/2012ApJ...761....6M} {761, 6}

\bibitem[\protect\citeauthoryear{{Morton}}{{Morton}}{2015}]{2015ascl.soft03010M}
{Morton} T.~D.,  2015, {isochrones: Stellar model grid package} (\mn@eprint
  {ascl} {1503.010})

\bibitem[\protect\citeauthoryear{{Morton}, {Bryson}, {Coughlin}, {Rowe},
  {Ravichandran}, {Petigura}, {Haas}  \& {Batalha}}{{Morton}
  et~al.}{2016}]{2016ApJ...822...86M}
{Morton} T.~D.,  {Bryson} S.~T.,  {Coughlin} J.~L.,  {Rowe} J.~F.,
  {Ravichandran} G.,  {Petigura} E.~A.,  {Haas} M.~R.,   {Batalha} N.~M.,
  2016, \mn@doi [\apj] {10.3847/0004-637X/822/2/86}, \href
  {https://ui.adsabs.harvard.edu/abs/2016ApJ...822...86M} {822, 86}

\bibitem[\protect\citeauthoryear{{Oliphant}}{{Oliphant}}{2007}]{2007CSE.....9c..10O}
{Oliphant} T.~E.,  2007, \mn@doi [Computing in Science and Engineering]
  {10.1109/MCSE.2007.58}, \href
  {https://ui.adsabs.harvard.edu/abs/2007CSE.....9c..10O} {9, 10}

\bibitem[\protect\citeauthoryear{{Paxton}, {Bildsten}, {Dotter}, {Herwig},
  {Lesaffre}  \& {Timmes}}{{Paxton} et~al.}{2011}]{2011ApJS..192....3P}
{Paxton} B.,  {Bildsten} L.,  {Dotter} A.,  {Herwig} F.,  {Lesaffre} P.,
  {Timmes} F.,  2011, \mn@doi [\apjs] {10.1088/0067-0049/192/1/3}, \href
  {https://ui.adsabs.harvard.edu/abs/2011ApJS..192....3P} {192, 3}

\bibitem[\protect\citeauthoryear{{Pecaut} \& {Mamajek}}{{Pecaut} \&
  {Mamajek}}{2013}]{2013ApJS..208....9P}
{Pecaut} M.~J.,  {Mamajek} E.~E.,  2013, \mn@doi [\apjs]
  {10.1088/0067-0049/208/1/9}, \href
  {https://ui.adsabs.harvard.edu/abs/2013ApJS..208....9P} {208, 9}

\bibitem[\protect\citeauthoryear{{Rauer} et~al.,}{{Rauer}
  et~al.}{2014}]{2014ExA....38..249R}
{Rauer} H.,  et~al., 2014, \mn@doi [Experimental Astronomy]
  {10.1007/s10686-014-9383-4}, \href
  {https://ui.adsabs.harvard.edu/abs/2014ExA....38..249R} {38, 249}

\bibitem[\protect\citeauthoryear{Reback et~al.,}{Reback
  et~al.}{2021}]{jeff-reback-2021-5501881}
Reback J.,  et~al., 2021, pandas-dev/pandas: Pandas 1.3.3,
  \mn@doi{10.5281/zenodo.5501881}, \url
  {https://doi.org/10.5281/zenodo.5501881}

\bibitem[\protect\citeauthoryear{{Ricker} et~al.,}{{Ricker}
  et~al.}{2014}]{2014SPIE.9143E..20R}
{Ricker} G.~R.,  et~al., 2014, in {Oschmann} Jacobus~M. J.,  {Clampin} M.,
  {Fazio} G.~G.,   {MacEwen} H.~A.,  eds,  Society of Photo-Optical
  Instrumentation Engineers (SPIE) Conference Series Vol. 9143, Space
  Telescopes and Instrumentation 2014: Optical, Infrared, and Millimeter Wave.
  p. 914320 (\mn@eprint {arXiv} {1406.0151}), \mn@doi{10.1117/12.2063489}

\bibitem[\protect\citeauthoryear{{Ricker} et~al.,}{{Ricker}
  et~al.}{2015}]{2015JATIS...1a4003R}
{Ricker} G.~R.,  et~al., 2015, \mn@doi [Journal of Astronomical Telescopes,
  Instruments, and Systems] {10.1117/1.JATIS.1.1.014003}, \href
  {https://ui.adsabs.harvard.edu/abs/2015JATIS...1a4003R} {1, 014003}

\bibitem[\protect\citeauthoryear{{Rowe} et~al.,}{{Rowe}
  et~al.}{2015}]{2015ApJS..217...16R}
{Rowe} J.~F.,  et~al., 2015, \mn@doi [\apjs] {10.1088/0067-0049/217/1/16},
  \href {https://ui.adsabs.harvard.edu/abs/2015ApJS..217...16R} {217, 16}

\bibitem[\protect\citeauthoryear{{Shporer} et~al.,}{{Shporer}
  et~al.}{2017}]{2017ApJ...847L..18S}
{Shporer} A.,  et~al., 2017, \mn@doi [\apjl] {10.3847/2041-8213/aa8bff}, \href
  {https://ui.adsabs.harvard.edu/abs/2017ApJ...847L..18S} {847, L18}

\bibitem[\protect\citeauthoryear{{Smith} et~al.,}{{Smith}
  et~al.}{2012}]{2012PASP..124.1000S}
{Smith} J.~C.,  et~al., 2012, \mn@doi [\pasp] {10.1086/667697}, \href
  {https://ui.adsabs.harvard.edu/abs/2012PASP..124.1000S} {124, 1000}

\bibitem[\protect\citeauthoryear{{Stassun} et~al.,}{{Stassun}
  et~al.}{2018}]{2018AJ....156..102S}
{Stassun} K.~G.,  et~al., 2018, \mn@doi [\aj] {10.3847/1538-3881/aad050}, \href
  {https://ui.adsabs.harvard.edu/abs/2018AJ....156..102S} {156, 102}

\bibitem[\protect\citeauthoryear{{Stassun} et~al.,}{{Stassun}
  et~al.}{2019}]{2019AJ....158..138S}
{Stassun} K.~G.,  et~al., 2019, \mn@doi [\aj] {10.3847/1538-3881/ab3467}, \href
  {https://ui.adsabs.harvard.edu/abs/2019AJ....158..138S} {158, 138}

\bibitem[\protect\citeauthoryear{{Stumpe}, {Smith}, {Catanzarite}, {Van Cleve},
  {Jenkins}, {Twicken}  \& {Girouard}}{{Stumpe}
  et~al.}{2014}]{2014PASP..126..100S}
{Stumpe} M.~C.,  {Smith} J.~C.,  {Catanzarite} J.~H.,  {Van Cleve} J.~E.,
  {Jenkins} J.~M.,  {Twicken} J.~D.,   {Girouard} F.~R.,  2014, \mn@doi [\pasp]
  {10.1086/674989}, \href
  {https://ui.adsabs.harvard.edu/abs/2014PASP..126..100S} {126, 100}

\bibitem[\protect\citeauthoryear{{Szentgyorgyi} \& {Fur{\'e}sz}}{{Szentgyorgyi}
  \& {Fur{\'e}sz}}{2007}]{2007RMxAC..28..129S}
{Szentgyorgyi} A.~H.,  {Fur{\'e}sz} G.,  2007, in {Kurtz} S.,  ed.,  Revista
  Mexicana de Astronomia y Astrofisica Conference Series Vol. 28, Revista
  Mexicana de Astronomia y Astrofisica Conference Series. pp 129--133

\bibitem[\protect\citeauthoryear{{Tala}, {Berdja}, {Jones}, {Vanzi}, {Ropert},
  {Flores}  \& {Viscasillas}}{{Tala} et~al.}{2014}]{2014SPIE.9147E..89T}
{Tala} M.,  {Berdja} A.,  {Jones} M.,  {Vanzi} L.,  {Ropert} S.,  {Flores} M.,
   {Viscasillas} C.,  2014, in {Ramsay} S.~K.,  {McLean} I.~S.,   {Takami} H.,
  eds,  Society of Photo-Optical Instrumentation Engineers (SPIE) Conference
  Series Vol. 9147, Ground-based and Airborne Instrumentation for Astronomy V.
  p. 914789, \mn@doi{10.1117/12.2056551}

\bibitem[\protect\citeauthoryear{{Tokovinin}, {Fischer}, {Bonati}, {Giguere},
  {Moore}, {Schwab}, {Spronck}  \& {Szymkowiak}}{{Tokovinin}
  et~al.}{2013}]{2013PASP..125.1336T}
{Tokovinin} A.,  {Fischer} D.~A.,  {Bonati} M.,  {Giguere} M.~J.,  {Moore} P.,
  {Schwab} C.,  {Spronck} J. F.~P.,   {Szymkowiak} A.,  2013, \mn@doi [\pasp]
  {10.1086/674012}, \href
  {https://ui.adsabs.harvard.edu/abs/2013PASP..125.1336T} {125, 1336}

\bibitem[\protect\citeauthoryear{{Van der Walt}, {Colbert}  \&
  {Varoquaux}}{{Van der Walt} et~al.}{2011}]{2011CSE....13b..22V}
{Van der Walt} S.,  {Colbert} S.~C.,   {Varoquaux} G.,  2011, \mn@doi
  [Computing in Science and Engineering] {10.1109/MCSE.2011.37}, \href
  {https://ui.adsabs.harvard.edu/abs/2011CSE....13b..22V} {13, 22}

\bibitem[\protect\citeauthoryear{{Whitford}}{{Whitford}}{1958}]{1958AJ.....63..201W}
{Whitford} A.~E.,  1958, \mn@doi [\aj] {10.1086/107725}, \href
  {https://ui.adsabs.harvard.edu/abs/1958AJ.....63..201W} {63, 201}

\bibitem[\protect\citeauthoryear{{Winn}}{{Winn}}{2010}]{2010arXiv1001.2010W}
{Winn} J.~N.,  2010, arXiv e-prints, \href
  {https://ui.adsabs.harvard.edu/abs/2010arXiv1001.2010W} {p. arXiv:1001.2010}

\bibitem[\protect\citeauthoryear{{Zhou} et~al.,}{{Zhou}
  et~al.}{2017}]{2017AJ....153..211Z}
{Zhou} G.,  et~al., 2017, \mn@doi [\aj] {10.3847/1538-3881/aa674a}, \href
  {https://ui.adsabs.harvard.edu/abs/2017AJ....153..211Z} {153, 211}

\bibitem[\protect\citeauthoryear{{Zhou} et~al.,}{{Zhou}
  et~al.}{2022}]{2022AJ....163..289Z}
{Zhou} G.,  et~al., 2022, \mn@doi [\aj] {10.3847/1538-3881/ac69e3}, \href
  {https://ui.adsabs.harvard.edu/abs/2022AJ....163..289Z} {163, 289}

\makeatother
\end{thebibliography}




\appendix

\section{Supplementary tables}

\begin{table*}
\caption{Statistically vetted candidates and validated planets (i.e., those with the suffix `b') orbiting solar-analogue stars.}
\label{tab:validated}
\begin{tabular}{lllllllllllll}
\hline
\hline
TOI     & Tmag $^a$ & Vmag    & Period     & R$_{p, \rm \textsc{vespa}}$ & R$_{p, \textit{TESS}}^i$ & R$_{p, \rm undil.}$ & Secth. $^b$ & Maxrad $^c$ & FPP $^d$    & Centr.      & On-target$^g$ & Priority\\ 
        &         &         & (day)      & ($R_{\earth}$)       & ($R_{\earth}$)     & ($R_{\earth}$)           &           & (arcsec)    &          & test & result   & \& obs.  \\ \hline
1689.01 & 6.3196  & 6.996 & 9.12381    & 2.781     & 2.109   & 2.108         & 0.00029   & 1.5    & 1.31e-05$^j$ & failed$^*$     & unclear$^k$    &    -\\
2545 b & 8.9059 & 9.521   & 7.994037   & 2.526     & 2.750   & 2.75          & 0.00033   & 1.5    & 0.00111$^j$  & passed    & confirmed      &  1: PRV\\
2569.01 & 11.1769 & 11.775  & 13.114774  & 8.551     & 10.489  & 13.235        & 0.00234   & 103    & 0.00874  & failed    & unclear      &  4: on-off\\
3353.01$^*$ & 8.7843  & 9.327   & 4.665774   & 2.840     & 2.831   & 2.834         & 0.00037   & 103    & 1.25e-13$^j$ & passed$^*$   & unclear      &  4: on-off\\
3353.02$^*$ & 8.7843  & 9.327   & 8.817565   & 2.468     & 2.464   & 2.477         & 0.00052   & 103    & 0.0$^j$      & passed$^*$   & unclear      &  4: on-off\\
3474.01 & 11.9672 & 12.531  & 3.878932   & 16.305    & 15.071  & 17.858        & 0.00239   & 82     & 8.02e-07 & failed    & unclear      &  4: on-off\\
3837.01 & 11.366  & 11.673  & 11.892894  & 13.496    & 12.036  & 12.11         & 0.00127   & 1.5    & 7.41e-13 & passed     & enhanced      &  2: HRI$^f$\\
3892.01 & 11.9956 & 12.607  & 4.581080   & 13.232    & 13.858  & 14.358        & 0.00179   & 82     & 1.01e-09 & failed    & unclear     &   4: on-off\\ 
4029.01 & 11.1718 & 11.554  & 5.884856   & 6.701     & 6.969   & 7.135         & 0.00106   & 103    & 0.00322  & passed    & enhanced      &  2: HRI\\
4361.01 & 8.6661  & 9.265   & 741.42559$^h$ & 2.931     & 2.971   & 2.971         &   9e-05     & 1.5    & 0.000626 & passed     & enhanced      &  2: HRI\\
4399 b & 7.7582  & 8.31    & 7.712121   & 3.310     & 3.208   & 3.207         & 0.00107   & 1.5    & 8.75e-06$^j$ & passed     & confirmed     &   1: PRV\\
4402.01 & 9.8429  & 10.286  & 3.698994   & 1.748     & 1.786   & 1.787         & 0.00012   & 93     & 0.00675  & failed$^*$    & unclear     &   4: on-off\\
4443.02 & 7.9147  & 8.493   & 10.313947  & 2.344     & 2.161   & 2.164         & 0.00020   & 92.5   & 1.21e-05$^j$ & passed    & confirmed     &  3: LPRV$^e$\\
4492.01 & 9.6311 & 10.324  & 4.433206   & 14.323    & 13.290  & 14.218        & 0.00075   & 103  & 1.74e-06 & failed    & unclear      &  4: on-off\\
4602.01 & 7.7746  & 8.32    & 3.980286   & 2.302     & 2.427   & 2.429              & 0.00014   & 82     & 0.0      & -    & unclear      &  4: on-off\ \\
4640.01 & 11.0771 & 11.63   & 2.685723   & 2.987     & 2.928   & 2.929         & 0.00038   & 1.5    & 0.000503 & passed     & enhanced      &  3: LPRV\\
4702.01 & 12.2469 & 12.877   & 3.121702   & 15.850     & 15.823   & 15.873         & 0.00098   & 1.5    & 0 & -     & enhanced      &  3: LPRV\\
4994.01 & 11.9545 & 12.652   & 21.492146   & 9.955     &  9.328  & 9.338         & 0.002817   & 1.5    & 2.64e-06 & passed     & enhanced     &  3: LPRV\\
5174 b & 10.6309 & 11.583   & 12.214286   & 5.346    &  5.343  & 5.351         & 0.00103   & 1.5    & 1.97e-04 & -    & confirmed      &  1: PRV\\
5210.01 & 11.4194 & 12.118   & 4.566131   & 13.341    &  12.228  & 12.827         & 0.00158   & 103    & 0 &  failed     & unclear      &  4: on-off\\
5238 b & 11.6370 & 12.214   & 4.872171   & 5.170    &  5.209  & 5.220         & 0.00268   & 82    & 1.43e-13 & passed     & confirmed      &  1: PRV\\
5398 b & 9.5806 & 10.059   & 10.590923   & 11.758    &  11.653  & 11.657         &  0.00232   & 1.5    & 3.26e-14 & -    & confirmed      & 1: PRV\\
5427.01 & 11.6590 & 12.140   & 5.237418   & 14.321    &  14.918  & 16.112         &  0.00306   & 82    & 7.44e-6 & failed    & unclear      &  4: on-off\\
\hline
\multicolumn{13}{l}{\textbf{ Notes.}}
\\
\multicolumn{13}{l}{$^a$ \textit{TESS} magnitude.}\\
\multicolumn{13}{l}{$^b$ Maximum secondary eclipse depth allowed.}\\
\multicolumn{13}{l}{$^c$ Exclusion radius within which FP scenarios are allowed.}\\
\multicolumn{13}{l}{$^d$ False Positive Probability.}\\
\multicolumn{13}{l}{$^e$ Low Precision Radial Velocity.}\\
\multicolumn{13}{l}{$^f$ High-resolution imaging.}\\
\multicolumn{13}{l}{$^g$ On-target probability. Full explanation in Section \ref{sec:vett-conf}.}\\
\multicolumn{13}{p{\textwidth}}{$^h$ TOI 4361.01 is a `duo-transit' candidate, i.e., a TOI with only two transits separated by about two years. Therefore, its period is not uniquely constrained but somewhat ambiguous. However, as further explained in Appendix \ref{appendix-b}, we have reasons to keep it as a vetted candidate regardless of its uncertain periodicity.}\\ 
\multicolumn{13}{p{\textwidth}}{$^i$ The \textit{TESS} planetary radius R$_{p, \textit{TESS}}$ has been calculated, in this work, from the transit depth available on the ExoFOP website and eq. 22 from \cite{2010arXiv1001.2010W}. We did the latter to maintain consistency with R$_{p, \rm \textsc{vespa}}$ \& R$_{p, \rm undil.}$, which were both calculated with the ExoFOP transit depth as a prior parameter. }\\ 
\multicolumn{13}{p{\textwidth}}{$^j$ TOI 3353.01 and TOI 3353.02 have high-resolution speckle imaging publicly available on the ExoFOP website (PI: Howell). Following \cite{2012ApJ...761....6M} and the \textsc{vespa} tutorial, we inserted the Gemini/Zorro contrast curves into the \textsc{vespa} input parameters. The final FPP of TOI 3353.01 slightly increases (reaching a value of 1e-6), while that of TOI 3353.02 remains unchanged. This result further confirms our statistical vetting. TOI 2545.01, TOI 4399.01, and TOI 4443.02 also have high-resolution speckle (or Adaptive Optics) imaging publicly available on the ExoFOP website (PI: Dressing, Howell, and Ciardi, respectively). The final FPP of TOI 2545.01 remains the same, while the FPP of every other TOIs decreases from two to seven orders of magnitude. Also in this case, our vettings are confirmed. This analysis allowed us to validate TOI 2545.01 and TOI 4399.01 and label them as TOI 2545 b and TOI 4399 b.}\\
\multicolumn{13}{p{\textwidth}}{$^k$ Companion detected at 0.08 arcseconds separation using high-resolution speckle imaging (Dr. Boris Safonov, from \url{exofop.ipac.caltech.edu}). New imaging data has been scheduled, and new analyses of existing data are also in progress (Dr. Boris Safonov, private communication). Additional information is available in Table \ref{tab:spectra}.}\\
\multicolumn{10}{l}{$^*$ Controversial.}\\
\end{tabular}
\end{table*}


\begin{table*}
\caption{Published spectra of the host stars of our targets. The column under `Total' demonstrates how many spectra in total are obtained by the same facility throughout the years.}
\label{tab:spectra}
\begin{tabular}{llllllll}
\hline
\hline
TOI     & \textit{V}    & Telescope & Instrument & Resolution & Spectral Range & Total & ExoFOP website's `Open Observing Notes'$^a$ \\ 
        &     (mag) &    & & & (\AA) && \\ \hline
1689.01 &  6.996 &  FLWO (1.5 m) & TRES & 44000 & 3850--9096  & 3 & Large RV offset. Potential composite spectrum.$^b$ \\
2545.01 &  9.521   & FLWO (1.5 m) & TRES & 44000 & 3850--9096  & 4 & False Positive scenarios ruled out.\\
2545.01 &  9.521   & SMARTS (1.5 m) & CHIRON & 80000 & 4500--8900  & 2 & --\\
2545.01 &  9.521   & ESO 1m telescope & FIDEOS & 43000 & 4200--8000  & 2 & --\\

2569.01 &  11.775  & FLWO (1.5 m) & TRES & 44000 & 3850--9096  & 2 & RV offset out-of-phase probably not significant. \\
3353.01 &  9.327  & SMARTS (1.5 m) & CHIRON & 80000 & 4500--8900  & 1 & --\\
3837.01 &  11.673  & FLWO (1.5 m) & TRES & 44000 & 3850--9096  & 3 & False Positive scenarios ruled out.\\
3892.01 &  12.607  &FLWO (1.5 m) & TRES & 44000 & 3850--9096  & 2 & False Positive scenarios ruled out.\\ 
4029.01 &  11.554  & FLWO (1.5 m) & TRES & 44000 & 3850--9096  & 2 & False Positive scenarios ruled out.\\
4361.01 &  9.265   & FLWO (1.5 m) & TRES & 44000 & 3850--9096  & 4 & False Positive scenarios ruled out.\\
4361.01 &  9.265   & SMARTS (1.5 m) & CHIRON & 80000 & 4500--8900  & 1 & --\\

4399.01 &  8.31    & SMARTS (1.5 m) & CHIRON & 80000 & 4500--8900  & 11 & False Positive scenarios ruled out.$^c$\\
4402.01 &  10.286  & SMARTS (1.5 m) & CHIRON & 80000 & 4500--8900  & 1 & -- \\
4443.01 &  8.493   & FLWO (1.5 m) & TRES & 44000 & 3850--9096  & 2 & RV offset out-of-phase. More observations needed.$^d$\\
4492.01 &  10.324  & FLWO (1.5 m) & TRES & 44000 & 3850--9096  & 2 & False Positive scenarios ruled out.\\
4602.01 &  8.32   & FLWO (1.5 m) & TRES & 44000 & 3850--9096  & 2 & False Positive scenarios ruled out.\\
4640.01 &  11.63   & FLWO (1.5 m) & TRES & 44000 & 3850--9096  & 2 & RV offset out-of-phase. More observations needed.$^d$\\
4702.01 &  12.877   & FLWO (1.5 m) & TRES & 44000 & 3850--9096  & 2 & More observations needed.\\
5174.01 &  11.583   & FLWO (1.5 m) & TRES & 44000 & 3850--9096  & 2 & False Positive scenarios ruled out.\\
5210.01 &  12.118   & FLWO (1.5 m) & TRES & 44000 & 3850--9096  & 2 & More observations needed.\\
5238.01 &  12.214   & FLWO (1.5 m) & TRES & 44000 & 3850--9096  & 2 & False Positive scenarios ruled out.\\
5398.01 &  10.059   & FLWO (1.5 m) & TRES & 44000 & 3850--9096  & 11 & False Positive scenarios ruled out.$^d$\\
\hline
\multicolumn{8}{l}{\textbf{ Notes.}}
\\
\multicolumn{8}{l}{$^a$ Summary of the `Open Observing Notes' publicly available on the ExoFOP website.} \\
\multicolumn{8}{p{\textwidth}}{$^b$ `The new TRES observation is very strong and is shifted by about 3 km/s compared to the first two TRES observations more than a year ago.  Moreover, there is more line broadening, hinting at a composite spectrum.  This is not a good target for PRV work or atmospheric characterization.  No more TRES recon spectra are needed.' (Dr. David Latham, from \url{exofop.ipac.caltech.edu}).} \\
\multicolumn{8}{l}{$^c$ Data publicly available on the ExoFOP website and further analysed by \cite{2022AJ....163..289Z}.} \\
\multicolumn{8}{p{\textwidth}}{$^d$ This conclusion comes from the `Open Observing Notes' and private communication with Dr. David Latham.} \\

\end{tabular}
\end{table*}

\section{Duo-transit candidate vetting}
\label{appendix-b}

\begin{figure}
	\includegraphics[width=\columnwidth]{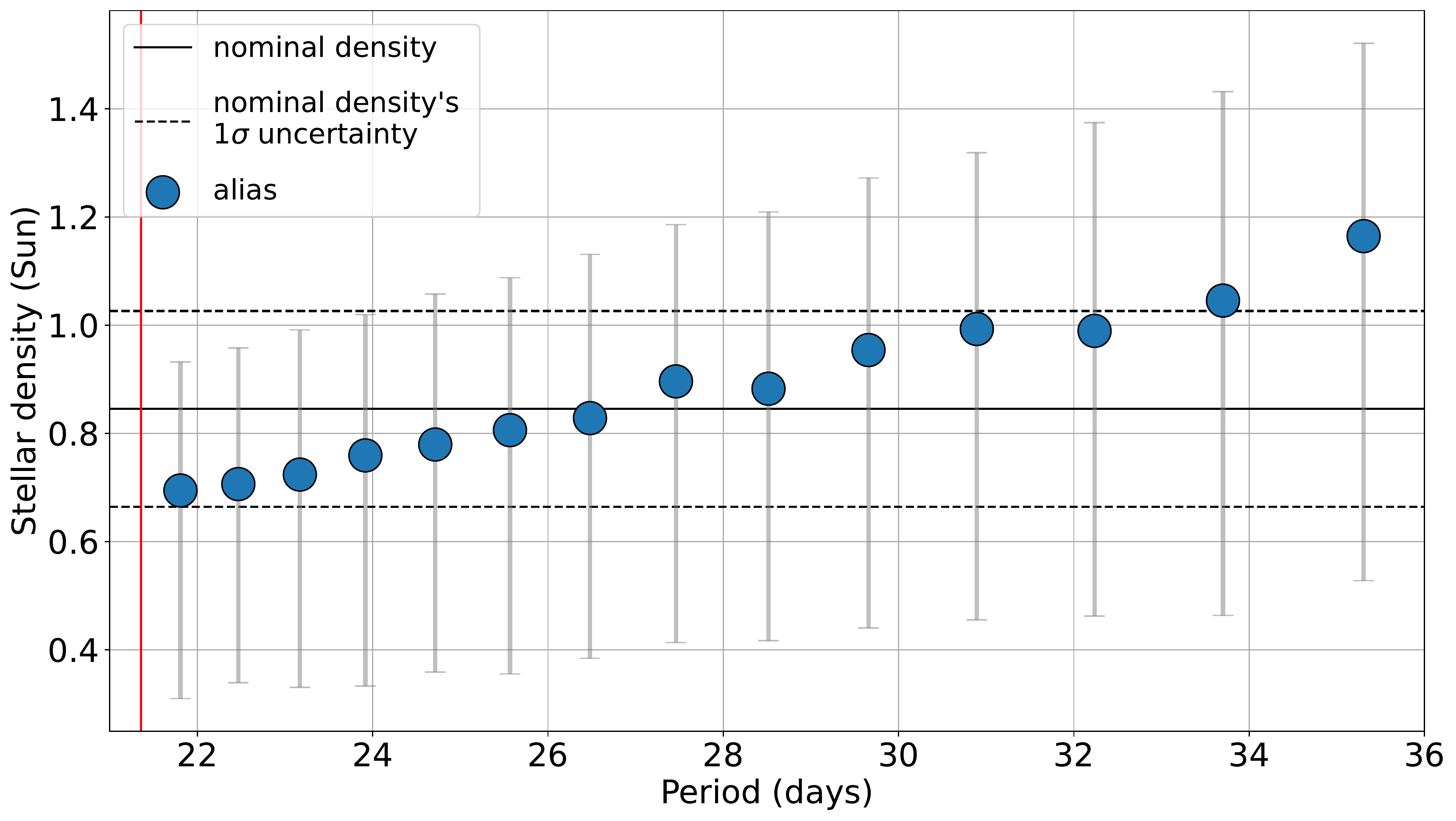}
    \caption{Stellar density as a function of TOI 4361.01 period aliases. On the x-axis, we represent the period value of a specific TOI 4361.01 alias, while on the y-axis, its host star's stellar density estimated with the MCMC simulation. Each blue dot represents a different TOI 4361.01 period alias, while the solid and dashed black lines show the nominal density and its uncertainty, respectively. The vertical red line indicates the minimum periodicity value.}
    \label{fig:mcmc}
\end{figure}

The \textit{TESS} candidate TOI 4361.01 is one of our vetted candidates with the second highest priority mark, which means that following our procedure, we demonstrated that it is ready to be analysed with high-resolution imaging and subsequently confirmed through high-precision radial velocity observations. However, there is a large gap in the \textit{TESS} data that causes its period $P$ of $\approx$741 days to be ambiguous. The absence of a period uniquely constrained induces us to be careful and requires further analysis to confirm its vetting. Therefore, we performed the following:
\begin{enumerate}[leftmargin=*]
    \item We took into account all possible TOI 4361.01 period aliases, and modelled the \textit{TESS} light curve using the \textsc{pycheops} code, to extrapolate the host star's stellar density $\rho$ from the transit signal (see Sec. \ref{sec:on-off}). We then ran an MCMC simulation to better estimate the value and the uncertainty of $\rho$;
    \item we considered all the aliases whose extrapolated $\rho$ has a physical result and performed, again, the \textsc{vespa} analysis considering the new period.
\end{enumerate} In Figure \ref{fig:mcmc}, we illustrate the results. In particular, we show the stellar density $\rho$ coming from the possible aliases as a function of the period aliases. We added the nominal stellar density available on the ExoFOP website for comparison. The lower limit on the orbital period comes from the ephemeris window covered by the \textit{TESS} mission during sectors 8 and 35, while the upper limit comes from the light curve modelling, where we fixed the transit duration as reported on the ExoFOP website. Specifically, when a period alias is $\gtrapprox  35.5$ days, the impact parameter $b$ (calculating following Eq. 7 from \citealt{2010arXiv1001.2010W} and considering a circular orbit) becomes $> 1$ (not physically allowed). These two limits constrain the period aliases to be within [$P$/34, $P$/21]. The assumption of a circular orbit avoids the treatment of eccentricity in the evaluation of impact parameter and the consequent degeneracy in the \textsc{pycheops} transit modelling. Although the latter assumption is simplistic, the results of exoplanet population studies, especially for low-mass, sub-Neptunes planets, favour low eccentricities values \citep{2011arXiv1109.2497M, 2012MNRAS.425..757K, 2013MNRAS.434L..51K}. From the results of this figure, we conclude that whilst this approach could yield a unique orbital period, we do not have the sampling in the transit ingress/egress to accurately say what it is. However, we emphasise that the unique orbital period estimation is not the central goal of this appendix and this work. 

Following the result of our analysis, we performed the \textsc{vespa} analysis of every survived TOI 4361.01 period alias. Moreover, for the sake of completeness, we also considered some period aliases ($P$/2, $P$/4, $P$/8, and $P$/16) outside the aforementioned limits. In conclusion, we have found the following:
\begin{itemize}[leftmargin=*]
    \item every TOI 4361.01 period alias shows a FPP $<$ 1\%;
    \item the shorter the period, the greater the planetary probability (and the lower the FPP). 
\end{itemize} We hence confirm our statistical analysis and keep TOI 4361.01 as one of our best vetted candidates.



\bsp	
\label{lastpage}
\end{document}